\documentclass[12pt,review]{elsarticle}
\usepackage[top=0.8in, bottom=0.8in, left=0.9in, right=0.9in]{geometry}

\graphicspath{ {figures/} }
\usepackage{graphicx}
\usepackage{subcaption} % 替代过时的 subfigure 包
\usepackage{tabularx}
\usepackage{color}
\usepackage{chngcntr}

\usepackage{amsmath}
\usepackage{amssymb}
\usepackage{amsthm}

\usepackage{comment}
\usepackage[unicode]{hyperref}
\usepackage{tabularx}
\usepackage{multirow} 
\usepackage{booktabs}

\journal{Elsevier}
\begin{document}

\begin{frontmatter}



\title{Vectorial discrete unified gas kinetic scheme for continuum compressible flows}


\author[inst1]{Yinghao Ma}
\author[inst2]{Ziyang Xin}
\author[inst3]{Zhaoli Guo\corref{cor1}}

\affiliation[inst1]{organization={School of Mathematics and Statistics},%Department and Organization
            addressline={Huazhong University of Science and Technology}, 
            city={Wuhan},
            postcode={430074}, 
            %%state={State One},
            country={China}}
\affiliation[inst2]{organization={State Key Laboratory of Coal Combustion, School of Energy and Power Engineering},%Department and Organization
            addressline={Huazhong University of Science and Technology}, 
            city={Wuhan},
            postcode={430074}, 
            %%state={State One},
            country={China}}
% \affiliation[inst2]{organization={Department of Mechanical Engineering},%Department and Organization
%             addressline={National University of Singapore}, 
%             city={10 Kent Ridge Crescent},
%             postcode={119260}, 
%             %%state={State One},
%             country={Singapore}}
\affiliation[inst3]{organization={Institute of Interdisciplinary Research for Mathematics and Applied Science},%Department and Organization
            addressline={Huazhong University of Science and Technology}, 
            city={Wuhan},
            postcode={430074}, 
            %%state={State One},
            country={China}}
\cortext[cor1]{Corresponding author. 
Email address: zlguo@hust.edu.cn
}

\begin{abstract}

A vectorial discrete unified gas kinetic scheme (V-DUGKS) is proposed for continuum compressible flow simulations. In the present vectorial kinetic framework, mass, momentum, and total energy are described by distinct yet coupled distribution functions, and separate relaxation processes are introduced for momentum and energy transport, allowing the Prandtl number to be independently prescribed. Based on Hermite expansion and Gauss--Hermite quadrature, all equilibrium distribution functions are truncated at the second-order Hermite level. Since only second-order velocity moments are required to recover the compressible Navier--Stokes (NS) equations, fourth-order Gauss--Hermite quadrature is sufficient for exact moment evaluation. The elimination of third-order Hermite contributions enables compact discrete velocity sets (DVSs) to be uniformly adopted for all distribution functions, thereby reducing the sensitivity to the numerical reference temperature and improving both stability and efficiency compared with the previous scalar DUGKS for continuum compressible flows. The proposed scheme is implemented within a finite-volume formulation with characteristic-based flux evaluation and trapezoidal time integration of the collision term. A series of numerical tests, including one-dimensional shock-tube problems, the Shu--Osher problem, a two-dimensional Riemann problem, and three-dimensional Taylor--Green vortex flows at various Mach numbers, are conducted to assess the accuracy, stability, and efficiency of the scheme. In particular, for the three-dimensional Taylor--Green vortex, the V-DUGKS achieves a speed-up factor of approximately $2.2\text{--}2.9$ over the scalar DUGKS under the same CFL constraint, owing to its larger allowable time step and simplified equilibrium formulation. Overall, the results demonstrate that the proposed V-DUGKS provides a robust and efficient kinetic framework for continuum compressible flow simulations.

\end{abstract}




\end{frontmatter}

\section{INTRODUCTION}
Vector--type BGK (Bhatnagar--Gross--Krook) models
have been proposed~\cite{bouchut1999construction,DELLAR200295,bouchut2003entropy} as extensions of classical scalar BGK models~\cite{BGK1954}, in which
the distribution function is vector--valued and its components are directly
associated with macroscopic conserved variables. Such vectorial kinetic models
provide a flexible theoretical framework for the approximation of systems of
conservation laws and form the kinetic basis of various vectorial numerical
schemes.
Building upon these vectorial kinetic models, extensive efforts have been
devoted to the theoretical and numerical analysis of the resulting schemes.
In particular, their consistency
\cite{Roberta2019,ZHAO2024115677}, stability
\cite{Guillon2024,ZHAO20202764}, and entropy--related properties
\cite{BIANCHINI2019280,Dubois2014} have been systematically investigated, leading to the
design of numerically robust schemes within the collide--and--stream framework
\cite{zhaojin2021,Zhao2020}. These studies clarified the stability mechanisms of vectorial formulations and demonstrated that, compared with their scalar counterparts, vectorial schemes can achieve stable and accurate solutions with fewer discrete velocities, thereby improving computational efficiency.

By applying the standard collide--and--stream discretization to vectorial
kinetic models, one obtains a class of numerical methods commonly referred to as
vectorial lattice Boltzmann methods (VLBM). Over the past two decades, VLBM has
been primarily developed and applied as a numerical framework for incompressible
flows~\cite{ZHAO2024115677,zhaojin2021,ZHAO2020123901,Chai_2022,BATY2023127667}.
Under low--Mach--number, VLBM--type schemes have been shown
to provide consistent approximations of the incompressible Navier--Stokes (NS)
equations, while retaining favorable numerical properties such as robustness and
compact discrete velocity sets. Consequently, VLBM has been widely recognized as
an effective alternative to scalar LBM for multi--dimensional
incompressible flow simulations.
More recently, several works have attempted to extend the VLBM framework beyond the
incompressible regime to compressible flow problems governed by the Euler
equations. These studies mainly focus on the robust capture of strong
discontinuities and development of structure-preserving schemes with desirable properties, such as
positivity, convexity properties and entropy-related stability~\cite{Wissocq2025,GRAILLE201474,Coulette2020,Dubois2014,caetano2019result,bellotti2025fourth}. However, such extensions
remain largely restricted to inviscid Euler systems, and the application of
VLBM to fully compressible NS equations is still very limited.

Among various kinetic solvers, the discrete unified gas kinetic scheme (DUGKS)~\cite{Guo2013DUGKS,Guo2015DUGKS} in finite-volume formulation, which was originally designed for flows covering multiple regimes, has been successfully extended to the simulation of continuum compressible flows governed by the NS equations~\cite{Wenxin2021,GUO2023Compressible,qiyiming2022}.
Recent developments in unified gas-kinetic modeling have further highlighted the connection between kinetic and hydrodynamic descriptions, from the Boltzmann equation to the NS limit~\cite{guo2026unified}.
By coupling particle transport and collision effects in the flux reconstruction,
DUGKS achieves low numerical dissipation and good stability.
However, as a discrete velocity method, the numerical stability and efficiency of compressible DUGKS strongly depend on the underlying kinetic model and DVS, especially at moderate to high Mach numbers.
This has motivated a series of recent efforts to enhance the numerical stability
of DUGKS for compressible flows, including regularization techniques and
improvements in the treatment of the collision term and flux reconstruction
\cite{qi2023regularized,qi2024regularized}.
Motivated by these studies, we consider a vectorial kinetic model in which
the macroscopic conserved variables, including mass, momentum, and total energy,
are represented by distinct distribution functions.
Each distribution function evolves according to its own discrete-velocity equation
% lattice Boltzmann equation (LBE) 
and is responsible for the transport of a specific physical
quantity, while their coupled dynamics at the kinetic level enables a consistent
description of transport and dissipation processes and ensures the recovery of
the fully compressible NS equations.
Based on this vectorial kinetic formulation, we develop a V-DUGKS for continuum compressible flows, extending the application of vectorial kinetic models from inviscid Euler systems to the compressible NS equations within the DUGKS framework, which combines the advantages of vectorial kinetic models with the unified finite-volume flux reconstruction of DUGKS.

One of the key issues in the construction of the proposed V-DUGKS lies in the
selection of appropriate DVS.
Within the framework of Hermite expansion and Gauss--Hermite quadrature~\cite{SHAN_YUAN_CHEN_2006}, the
discrete velocities must be chosen such that the velocity moments required for
recovering the compressible NS equations can be evaluated exactly.
In this work, the notation D$n$V$b$--A$x$ is adopted to denote a discrete velocity
model, where $n$ represents the spatial dimension, $b$ is the number of discrete
velocity points, and $x$ indicates that the corresponding Gauss--Hermite
quadrature achieves $x$-th order orthogonal accuracy.
In early kinetic schemes for compressible flows, at least eighth--order Gauss--Hermite quadrature was typically required to ensure the correct recovery of macroscopic equations, leading to the use of large DVSs such as D3V77--A9
\cite{Wenxin2021} and D3V45--A9 \cite{WILDE2021101355}. 
Later, by introducing a double distribution
function (DDF) approach, it was shown in \cite{GUO2023Compressible} that the transport of different macroscopic quantities can be decoupled, so that the two distribution functions only need to satisfy fourth-- and sixth--order Gauss--Hermite quadrature, respectively. As a result, more compact DVSs, including
D3V13--A5 and D3V27--A7 can be employed.
In the present vectorial kinetic model, each conserved variable is represented
by a structurally independent distribution function, and only second--order
velocity moments are involved in the recovery of the macroscopic equations.
Therefore, fourth--order Gauss--Hermite quadrature is sufficient to accurately
recover the compressible NS equations, since all equilibrium
distribution functions in the present model are constructed using a
second--order Hermite expansion. Based on this consideration,
the D3V13--A5 set is uniformly adopted for all distribution
functions in this work, which significantly reduces the computational cost while
maintaining the desired numerical accuracy.

The remainder of this paper is organized as follows.
After introducing the vectorial kinetic model and the corresponding macroscopic compressible NS model
(Sec.~\ref{sec:2_1}), the Hermite expansion of the equilibrium distribution functions and the construction of the DVSs based on Gauss--Hermite quadrature are presented in Sec.~\ref{sec:2_2}.
The numerical realization and analysis of the proposed V-DUGKS are presented in
Sec.~\ref{sec:3}.
In Sec.~\ref{sec:4}, a series of numerical experiments for compressible flows are
performed to assess the numerical accuracy, stability, and computational efficiency of the proposed V-DUGKS, covering both one-dimensional shock-related benchmarks and multi-dimensional flow problems.
Finally, Sec.~\ref{sec:5} summarizes the main results of this work.

\section{VECTORIAL KINETIC MODEL FOR COMPRESSIBLE NAVIER--STOKES FLOWS}\label{sec:2}
\subsection{Vectorial kinetic model}\label{sec:2_1}
First, we consider the BGK model for a rarefied gas, 
% The evolution of the distribution function $f=f(\boldsymbol{x},\boldsymbol{\xi},\boldsymbol{\eta},\boldsymbol{\zeta},t)$ 
which is governed by
\begin{equation} \label{BGKeq}
\partial_t f + \boldsymbol{\xi}\cdot\nabla f
= -\frac{1}{\tau}\left(f-f^{(eq)}\right),
\end{equation}
where $f=f(\boldsymbol{x},\boldsymbol{\xi},\boldsymbol{\eta},\boldsymbol{\zeta},t)$  is the distribution function, $\boldsymbol{x}\in\mathbb{R}^D$ and $t$ denote the physical space and time, 
respectively. The particle velocity is decomposed into three parts: 
$\boldsymbol{\xi}=(\xi_1,\ldots,\xi_D)$ represents the translational velocity 
components associated with spatial transport, while 
$\boldsymbol{\eta}=(\xi_{D+1},\ldots,\xi_3)$, and 
$\boldsymbol{\zeta}$ is a vector of length $K$ denoting degrees of internal
freedom of gas molecules. Note that $\boldsymbol{\eta}$ and $\boldsymbol{\zeta}$ do not directly 
participate in particle advection~\cite{BGKmodel1993}. The equilibrium distribution function $f^{(eq)}$ is assumed to be a Maxwellian,
% that depends on the macroscopic density $\rho$, flow velocity $\boldsymbol{u}$, 
% and temperature $T$. It is written as
\begin{equation}
f^{(eq)}=
\frac{\rho}{(2\pi RT)^{(3+K)/2}}
\exp\!\left[-\frac{c^2+\eta^2+\zeta^2}{2RT}\right],
\end{equation}
where $\rho$, $\boldsymbol{u}$, and $T$ are the macroscopic density, velocity, and temperature, respectively. $R$ denotes the gas constant and 
$\boldsymbol{c}=\boldsymbol{\xi}-\boldsymbol{u}$ is the peculiar velocity.
The macroscopic variables are obtained as velocity moments of the distribution 
function,
\begin{equation}
\label{eq:macro_moments}
\begin{pmatrix}
\rho \\
\rho \boldsymbol{u} \\[6pt]
\dfrac{(3+K)\rho R T}{2}
\end{pmatrix}
=
\begin{pmatrix}
\displaystyle \int f\, d\boldsymbol{\xi}\, d\boldsymbol{\eta}\, d\boldsymbol{\zeta} \\[8pt]
\displaystyle \int \boldsymbol{\xi}\, f\, d\boldsymbol{\xi}\, d\boldsymbol{\eta}\, d\boldsymbol{\zeta} \\[8pt]
\displaystyle \int \frac{c^2+\eta^2+\zeta^2}{2}
f\, d\boldsymbol{\xi}\, d\boldsymbol{\eta}\, d\boldsymbol{\zeta}
\end{pmatrix}.
\end{equation}
% \begin{subequations}
% \begin{align}
% \rho &= \int f\, d\boldsymbol{\xi}\, d\boldsymbol{\eta}\, d\boldsymbol{\zeta}, \\
% \rho\boldsymbol{u} &= \int \boldsymbol{\xi}\, f\, d\boldsymbol{\xi}\, d\boldsymbol{\eta}\, d\boldsymbol{\zeta},
% \end{align}
% \end{subequations}
% while the temperature is related to the total kinetic energy of molecular motion,
% \begin{equation}
% \frac{3+K}{2}\rho RT
% = \int \frac{1}{2}\left(c^2+\eta^2+\zeta^2\right)
% f\, d\boldsymbol{\xi}\, d\boldsymbol{\eta}\, d\boldsymbol{\zeta}.
% \end{equation}
The total energy $E$ is defined as
\begin{equation}
\rho E = \rho e+\tfrac{1}{2}\rho u^2
= \int \frac{1}{2}\left(\xi^2+\eta^2+\zeta^2\right)
f\, d\boldsymbol{\xi}\, d\boldsymbol{\eta}\, d\boldsymbol{\zeta},
\end{equation}
with the internal energy $e=c_v T$ and $c_v=(3+K)R/2$ denotes the specific heat at constant volume.
For later use, the pressure tensor $\boldsymbol{P}$ and heat flux $\boldsymbol{q}$ 
are expressed as
\begin{equation}
\boldsymbol{P}=\int \boldsymbol{c}\boldsymbol{c}\, f\,
d\boldsymbol{\xi}\, d\boldsymbol{\eta}\, d\boldsymbol{\zeta},
\qquad
\boldsymbol{q}=\int \frac{1}{2}\left(c^2+\eta^2+\zeta^2\right)
\boldsymbol{c}\, f\,
d\boldsymbol{\xi}\, d\boldsymbol{\eta}\, d\boldsymbol{\zeta}.
\end{equation}

It is seen that the macroscopic variables are represented as velocity
moments of the distribution function in the classical scalar kinetic model.
% In conventional scalar formulations, the
% flow velocity is evaluated from a first-order moment, which increases the
% computational complexity of moment evaluations. 
To obtain a more compact and
structured representation, we propose a vectorial kinetic formulation in
which different macroscopic quantities are represented by different distribution
functions. 
% With this formulation, the flow velocity can be recovered from a
% zeroth-order moment of a vector-valued distribution function.
% Accordingly, the following reduced distribution functions are defined:
Specifically, we define the following distribution functions from the scalar one,
\begin{equation}
\label{vec-distribution}
\begin{pmatrix}
f_0(\boldsymbol{x}, \boldsymbol{\xi}, t) \\[6pt]
\boldsymbol{f}(\boldsymbol{x}, \boldsymbol{\xi}, t) \\[6pt]
f_{D+1}(\boldsymbol{x}, \boldsymbol{\xi}, t)
\end{pmatrix}
=
\begin{pmatrix}
\displaystyle \int f(\boldsymbol{x}, \boldsymbol{\xi}, \boldsymbol{\eta}, \boldsymbol{\zeta}, t)\,
{d}\boldsymbol{\eta}\, {d}\boldsymbol{\zeta} \\[10pt]
\displaystyle \int \boldsymbol{\xi}\,
f(\boldsymbol{x}, \boldsymbol{\xi}, \boldsymbol{\eta}, \boldsymbol{\zeta}, t)\,
{d}\boldsymbol{\eta}\, {d}\boldsymbol{\zeta} \\[10pt]
\displaystyle \int \frac{\xi^{2} + \eta^{2} + \zeta^{2}}{2}\,
f(\boldsymbol{x}, \boldsymbol{\xi}, \boldsymbol{\eta}, \boldsymbol{\zeta}, t)\,
{d}\boldsymbol{\eta}\, {d}\boldsymbol{\zeta}
\end{pmatrix}.
\end{equation}
Here, $f_0$ represents the density distribution function, $\boldsymbol{f}$ is
the momentum distribution function with $D$ components corresponding to the
spatial dimensions, and $f_{D+1}$ denotes the total energy distribution
function.

% In the kinetic formulation, macroscopic variables are obtained from velocity 
% moments of the distribution function. In particular, the flow velocity is usually 
% evaluated from a first-order moment, which increases the complexity of moment 
% calculations. To obtain a more compact representation, we introduce a 
% vectorial kinetic model in which different macroscopic quantities 
% are carried by different distribution functions. With this formulation, the flow 
% velocity can be recovered from a zeroth-order moment of a vector-valued 
% distribution function. Accordingly, the following reduced distribution functions are defined:
% \begin{equation} \label{vec-distribution}
% \begin{aligned}
% f_0(\boldsymbol{x}, \boldsymbol{\xi}, t)
% &= \int f(\boldsymbol{x}, \boldsymbol{\xi}, \eta, \zeta, t)\,
%    \mathrm{d}\eta\, \mathrm{d}\zeta, \\
% \boldsymbol{f}(\boldsymbol{x}, \boldsymbol{\xi}, t)
% &= \int \boldsymbol{\xi} f(\boldsymbol{x}, \boldsymbol{\xi}, \eta, \zeta, t)\,
%    \mathrm{d}\eta\, \mathrm{d}\zeta, \\
% f_{D+1}(\boldsymbol{x}, \boldsymbol{\xi}, t)
% &= \int \frac{\xi^{2} + \eta^{2} + \zeta^{2}}{2}\,
%    f(\boldsymbol{x}, \boldsymbol{\xi}, \eta, \zeta, t)\,
%    \mathrm{d}\eta\, \mathrm{d}\zeta .
% \end{aligned}
% \end{equation}

Based on Eq.~\eqref{vec-distribution}, a vectorial kinetic model can be constructed from the BGK equation~\eqref{BGKeq}, in which the collision process associated with different macroscopic quantities is allowed to relax with distinct relaxation times,
% The resulting vectorial kinetic equations governing the evolution 
% of the reduced distribution functions are given by
\begin{subequations}\label{vec-BGK}
\begin{align} 
\partial_t f_{\alpha} + \boldsymbol{\xi} \cdot \nabla f_{\alpha}
    &= - \frac{1}{\tau_f} \left( f_{\alpha} - f_{\alpha}^{(eq)} \right),
    \qquad \alpha = 0,1,\ldots,D, \label{BGK_a} \\[6pt]
\partial_t f_{D+1} + \boldsymbol{\xi} \cdot \nabla f_{D+1}
    &= - \frac{1}{\tau_e} \left( f_{D+1} - f_{D+1}^{(eq)} \right)
       + \frac{1}{\tau_{ef}} R,  \label{BGK_b}
\end{align}
\end{subequations}
where the coupling term $R=\sum_{\alpha=1}^{D}(f_\alpha-f_\alpha^{(eq)})u_\alpha$.
% In the present vectorial kinetic model, the relaxation processes 
% associated with momentum and energy are characterized by different time scales. 
The momentum relaxation time $\tau_f$ is related to the shear viscosity $\mu$ 
through
\begin{equation}
\tau_f = \frac{\mu}{p},
\end{equation}
where $p=\rho RT$ denotes the pressure. The energy relaxation time $\tau_e$ is further 
linked to $\tau_f$ via the Prandtl number $(\mathrm{Pr})$ as
\begin{equation}
\tau_e = \frac{\tau_f}{\mathrm{Pr}},
\end{equation}
% 2222222
% \begin{equation}
%     \partial_t f_{\alpha} + \boldsymbol{\xi} \cdot \nabla f_{\alpha}
%     = - \frac{1}{\tau_\alpha} \left( f_{\alpha} - f_{\alpha}^{(eq)} \right) 
%     +\frac{1}{\tau_{ef}}R_\alpha ,  \qquad \alpha=0,1,...,D+1,
% \end{equation}
% where $\tau_\alpha=\tau_f$ ($\alpha=0,..,D$) is the momentum relaxation time which is related to the dynamic viscosity through $\tau_f=\mu / p$, $p$ denotes the pressure. And $\tau_{D+1}=\tau_e$ is the energy relaxation time which is linked to $\tau_f$ via the Prandtl number $\mathrm{Pr}$ as $\tau_e=\tau/\mathrm{Pr}$.
% The coupling term $R_\alpha=0$ ($\alpha=0,...D$) and $R_{D+1}=\sum_{\alpha=1}^{D}(f_\alpha-f_\alpha^{(eq)})u_\alpha$.
and the coupling relaxation time $\tau_{ef}$ is defined by $\tau_{ef}^{-1} = \tau_e^{-1} - \tau_f^{-1}.$ With this formulation, the relaxation of the energy-related distribution function 
naturally accounts for both mechanical and internal energy contributions, each 
relaxing at its own time scale. As a consequence, Pr can be independently prescribed within the vectorial distribution functions.
The corresponding equilibrium distribution functions for the vectorial kinetic model are given by
% \begin{subequations}
% \begin{align}
% f_0^{(eq)}(\boldsymbol{\xi}; \rho, \boldsymbol{u}, T)
% &= \frac{\rho}{\left(2\pi R T\right)^{D/2}}
%    \exp\!\left[
%    - \frac{(\boldsymbol{\xi}-\boldsymbol{u})^{2}}{2 R T}
%    \right],  \\[6pt]
% \boldsymbol{f}^{(eq)}(\boldsymbol{\xi}; \rho, \boldsymbol{u}, T)
% &= \boldsymbol{\xi} f_0^{(eq)},  \\[6pt]
% f_{D+1}^{(eq)}(\boldsymbol{\xi}; \rho, \boldsymbol{u}, T)
% &= \frac{\xi^{2} + (3-D+K) R T}{2}\,
%    f_0^{(eq)}.
% \end{align}
% \end{subequations}
\begin{equation}
\begin{pmatrix}
f_0^{(eq)} \\[6pt]
\boldsymbol{f}^{(eq)} \\[6pt]
f_{D+1}^{(eq)}
\end{pmatrix}
=
\begin{pmatrix}
\displaystyle
\frac{\rho}{\left(2\pi R T\right)^{D/2}}
\exp\!\left[
- \frac{(\boldsymbol{\xi}-\boldsymbol{u})^{2}}{2 R T}
\right] \\[12pt]
\displaystyle
\boldsymbol{\xi}\, f_0^{(eq)} \\[12pt]
\displaystyle
\frac{\xi^{2} + (3-D+K) R T}{2}\,
f_0^{(eq)}
\end{pmatrix}.
\end{equation}

Accordingly, the macroscopic density $\rho$, velocity $\boldsymbol{u}$, and total energy $E$ are obtained from the zeroth-order moments of the 
corresponding distribution functions,
\begin{equation}\label{conservedeq}
\begin{pmatrix}
\rho \\[6pt]
\rho \boldsymbol{u} \\[6pt]
\rho E
\end{pmatrix}
=
\begin{pmatrix}
\displaystyle \int f_0\, d\boldsymbol{\xi} \\[6pt]
\displaystyle \int \boldsymbol{f}\, d\boldsymbol{\xi} \\[6pt]
\displaystyle \int f_{D+1}\, d\boldsymbol{\xi}
\end{pmatrix}
=
\begin{pmatrix}
\displaystyle \int f_0^{(eq)}\, d\boldsymbol{\xi} \\[6pt]
\displaystyle \int \boldsymbol{f}^{(eq)}\, d\boldsymbol{\xi} \\[6pt]
\displaystyle \int f_{D+1}^{(eq)}\, d\boldsymbol{\xi}
\end{pmatrix}.
\end{equation}
In addition, the pressure tensor and heat flux are defined through first-order moments of the vectorial distribution functions,
\begin{equation}
\boldsymbol{P}
= \int \boldsymbol{c}\, \boldsymbol{f} \, d{\boldsymbol{\xi}},
\qquad
\boldsymbol{q}
= \int \boldsymbol{c}\,(f_{D+1} - \boldsymbol{u} \cdot \boldsymbol{f})\, d{\boldsymbol{\xi}}
= \int \boldsymbol{c}\, f_{D+1}\, d{\boldsymbol{\xi}}
  - \boldsymbol{P}\cdot \boldsymbol{u}.
\end{equation}

% By applying the Chapman--Enskog (CE) expansion to the vectorial kinetic equations~\eqref{vec-BGK}, the following moments of the equilibrium distribution functions are required,
It is noted that
the equilibrium distribution functions satisfy the following moments,
\begin{subequations} \label{moments}
\begin{align}
\int \xi_i f_0^{(eq)} \, d\boldsymbol{\xi} &= \rho u_i, \quad
\int \xi_i \xi_j f_0^{(eq)} \, d\boldsymbol{\xi}
= p \delta_{ij} + \rho u_i u_j, \label{moments_a} \\[8pt]
% \Pi^{(0)}_{\alpha\beta} = 
\int \xi_i f_\alpha^{(eq)} \, d\boldsymbol{\xi} &= p \delta_{i \alpha} + \rho u_i u_\alpha, \quad
\int \xi_i \xi_j f_{\alpha}^{(eq)} \, d\boldsymbol{\xi}
= p (u_\alpha \delta_{ij}
      + u_i \delta_{\alpha j}
      + u_j \delta_{\alpha i})
  + \rho u_i u_j u_\alpha, \label{moments_b} \\[8pt]
% Q^{(0)}_{\alpha}= 
\int \xi_i f_{D+1}^{(eq)} \, d\boldsymbol{\xi} &= (p + \rho E) u_i, \quad
\int \xi_i \xi_j f_{D+1}^{(eq)} \, d\boldsymbol{\xi}
= p(RT + E)\,\delta_{ij} + (2p + \rho E) u_i u_j. \label{moments_c}
\end{align}
\end{subequations}
Then the compressible NS equations can be derived from Eq.~\eqref{vec-BGK} through the Chapman--Enskog (CE) analysis (See details in Appendix~A),
% recovered at the first order of the expansion (the detail is provided in Appendix~A),
\begin{subequations}\label{eq:NS}
\begin{align}
\partial_t \rho + \nabla \cdot (\rho \boldsymbol{u}) &= 0, \label{eq:NS_a} \\[6pt]
\partial_t (\rho \boldsymbol{u}) + \nabla \cdot (\rho \boldsymbol{u} \boldsymbol{u})
    &= - \nabla p + \nabla \cdot \boldsymbol{\sigma}, \label{eq:NS_b} \\[6pt]
\partial_t (\rho E) + \nabla \cdot \left[(p + \rho E)\boldsymbol{u}\right]
    &= \nabla \cdot (\kappa \nabla T) + \nabla \cdot (\boldsymbol{\sigma} \cdot \boldsymbol{u}), \label{eq:NS_c}
\end{align}
\end{subequations}
where $\kappa=\tau_e c_p p$ is the thermal conductivity, with
$c_p=(5+K)R/2$ being the specific heat at constant pressure and $\boldsymbol{\sigma}$ is the viscous stress tensor,
\begin{equation}
    \boldsymbol{\sigma}=2\mu\!\left[\boldsymbol{S}-\frac{1}{D}(\nabla\!\cdot\!\boldsymbol{u})\boldsymbol{I}\right]
+\mu_b(\nabla\!\cdot\!\boldsymbol{u})\boldsymbol{I}.
\end{equation}
Here $\boldsymbol{S}=\left[ \nabla \boldsymbol{u} + (\nabla \boldsymbol{u})^T \right]/2$ is the strain rate tensor, $\boldsymbol{I}$ is the identity tensor, $\mu=\tau_f p$ is the shear viscosity, and the bulk viscosity is $\mu_b = 2\mu(3-D+K)/(3+K)D$.

It is emphasized that the transport coefficients appearing in the above 
equations are determined by the separate relaxation times.
% introduced in the vectorial-distribution kinetic formulation. 
As a result, the present model 
recovers the compressible NS equations with independently controllable momentum and energy transport properties.

\subsection{Discrete velocity model}\label{sec:2_2}
To transform the vectorial kinetic model defined in the continuous velocity space into a computable numerical scheme, an appropriate discretization of the velocity space is required. In the present work, we adopt the approach proposed in Ref.~\cite{SHAN_YUAN_CHEN_2006}. First, the equilibrium distribution function $f_\alpha^{(eq)}$ ($\alpha = 0,1,..,D+1$) is expanded into their Hermite series,
\begin{equation}
\begin{aligned}
f_\alpha^{(eq)}(\hat{\boldsymbol{\xi}})
&\approx f_{\alpha,M}^{(eq)}(\hat{\boldsymbol{\xi}})
= \omega(\hat{\boldsymbol{\xi}})
  \sum_{n=0}^{M} \frac{1}{n!}\,
    \boldsymbol{a}_\alpha^{(n)}\, \boldsymbol{H}^{(n)}(\hat{\boldsymbol{\xi}}),
\end{aligned}
\end{equation}
where $\hat{\boldsymbol{\xi}}=\boldsymbol{\xi}/\sqrt{R T_{ref}}$ with $T_{ref}$ being a numerical reference temperature and the weight function is
$\omega(\hat{\boldsymbol{\xi}})=(2\pi)^{-D/2}\exp\!\left(-\tfrac{1}{2}|\hat{\boldsymbol{\xi}}|^2\right)$.
$\boldsymbol{H}^{(n)}(\hat{\boldsymbol{\xi}})$ $(n=0,1,\ldots)$ are the Hermite polynomials, 
$M$ is the truncation order, and $\boldsymbol{a}^{(n)}_{\alpha}$ are the corresponding expansion coefficients defined as 
\begin{equation}
\boldsymbol{a}_\alpha^{(n)}
= \int f_\alpha^{(eq)}(\hat{\boldsymbol{\xi}})
      \boldsymbol{H}^{(n)}(\hat{\boldsymbol{\xi}})\, d\hat{\boldsymbol{\xi}}.
\end{equation}

In order to recover the compressible NS equations, the moment conditions Eq.~\eqref{moments} must be satisfied at discrete level.
This suggests that the truncation order of the Hermite expansion of $f_\alpha^{(eq)}$ in \eqref{vec-BGK} is required to be $M=2$ at least. 
The subscript $M$ of the truncated distribution functions will be omitted hereafter for brevity. Then the explicit expansions can be written as
% \begin{align} \label{hermite1}
% f_{0}^{(eq)}(\hat{\boldsymbol{\xi}})
% &
% % = f_{0,2}^{(eq)}(\hat{\boldsymbol{\xi}})
%  = \rho\, \omega(\hat{\boldsymbol{\xi}})
%    \Bigl\{
%       1
%       + \hat{\boldsymbol{\xi}}\!\cdot\!\hat{\boldsymbol{u}}
%       + \tfrac{1}{2}
%         \bigl[
%             (\hat{\boldsymbol{\xi}}\!\cdot\!\hat{\boldsymbol{u}})^{2}
%             - \hat{u}^{2}
%             + (\theta - 1)(\hat{\xi}^{2} - D)
%         \bigr]
%    \Bigr\},
% \\[12pt] \label{hermite2}
% \boldsymbol{f}^{(eq)}(\hat{\boldsymbol{\xi}})
% &
% % = f_{\alpha,2}^{(eq)}(\hat{\boldsymbol{\xi}})
%  = \rho \, \theta \, \omega(\hat{\boldsymbol{\xi}})
%    \Bigl[
%      (\hat{\boldsymbol{\xi}}\!\cdot\!\hat{\boldsymbol{u}}) \boldsymbol{\xi}
%       + \boldsymbol{\xi} - \boldsymbol{u}
%    \Bigr]
%    + \boldsymbol{u} \, f_{0}^{(eq)}, 
% \\[12pt] \label{hermite3}
% f_{D+1}^{(eq)}(\hat{\boldsymbol{\xi}})
% &
% % = f_{D+1,2}^{(eq)}(\hat{\boldsymbol{\xi}})
%  = \omega(\hat{\boldsymbol{\xi}})\,
%     p\Bigl[
%         (\hat{\boldsymbol{\xi}}\!\cdot\!\hat{\boldsymbol{u}})
%         + (\hat{\boldsymbol{\xi}}\!\cdot\!\hat{\boldsymbol{u}})^{2}
%         - \hat{u}^{2}
%         + \tfrac{1}{2}\theta(\hat{\xi}^{2} - D)
%      \Bigr]
%      + E\, f_{0}^{(eq)} ,
% \end{align}
\begin{equation}\label{hermite}
    f_{\alpha}^{(eq)}(\hat{\boldsymbol{\xi}}) = 
    \begin{cases}
        \rho\, \omega(\hat{\boldsymbol{\xi}})
   \Bigl\{
      1
      + \hat{\boldsymbol{\xi}}\!\cdot\!\hat{\boldsymbol{u}}
      + \tfrac{1}{2}
        \bigl[
            (\hat{\boldsymbol{\xi}}\!\cdot\!\hat{\boldsymbol{u}})^{2}
            - \hat{u}^{2}
            + (\theta - 1)(\hat{\xi}^{2} - D)
        \bigr]
   \Bigr\}, \qquad \alpha = 0,
   \\[10pt]
   \rho \, \theta \, \omega(\hat{\boldsymbol{\xi}})
   \Bigl[
     (\hat{\boldsymbol{\xi}}\!\cdot\!\hat{\boldsymbol{u}}) {\xi_\alpha}
      + {\xi}_\alpha - {u}_\alpha
   \Bigr]
   + {u}_\alpha \, f_{0}^{(eq)}, \qquad \alpha = 1,...,D,
   \\[10pt]
   \omega(\hat{\boldsymbol{\xi}})\,
    p\Bigl[
        (\hat{\boldsymbol{\xi}}\!\cdot\!\hat{\boldsymbol{u}})
        + (\hat{\boldsymbol{\xi}}\!\cdot\!\hat{\boldsymbol{u}})^{2}
        - \hat{u}^{2}
        + \tfrac{1}{2}\theta(\hat{\xi}^{2} - D)
     \Bigr]
     + E\, f_{0}^{(eq)} , \qquad \alpha = D+1,
    \end{cases}
\end{equation}
where $\hat{\boldsymbol{u}}={\boldsymbol{u}}/\sqrt{RT_{ref}}$ and $\theta = T/T_{ref}.$

Next, the velocity discretization is carried out based on the Gauss--Hermite quadrature. Specifically, the following quadrature is employed,
\begin{align}
\int \hat{\boldsymbol{\xi}}^{m}\,
      f_{\alpha}^{(eq)}(\hat{\boldsymbol{\xi}})\,
      d\hat{\boldsymbol{\xi}}
&=
\int \hat{\boldsymbol{\xi}}^{m}\,
      \omega(\hat{\boldsymbol{\xi}})
      \bigl[\omega^{-1}(\hat{\boldsymbol{\xi}})\,
            f_{\alpha}^{(eq)}(\hat{\boldsymbol{\xi}})\bigr]\,
      d\hat{\boldsymbol{\xi}} \notag \\
&=
\sum_{k} w_{k}\,
      \hat{\boldsymbol{\xi}}_{k}^{m}
      \bigl[\omega^{-1}(\hat{\boldsymbol{\xi}}_{k})\,
            f_{\alpha}^{(eq)}(\hat{\boldsymbol{\xi}}_{k})\bigr],
\qquad
m = 0,1,2 ,    
\end{align}
where $\hat{\boldsymbol{\xi}}_{k}$ and $w_{k}$ are the corresponding abscissae and weights of the Gauss--Hermite quadrature, respectively. Since $\omega^{-1}(\hat{\boldsymbol{\xi}}_{k})\,f_{\alpha}^{(eq)}(\hat{\boldsymbol{\xi}}_{k})$ is a second-order polynomial in $\hat{\boldsymbol{\xi}_k}$, 
a quadrature rule of at least fourth-order accuracy is required to satisfy the moment conditions Eq.~\eqref{moments}. For example, the D3V13--A5 DVS listed in Table~\ref{tab:DVS} can be adopted for all equilibrium distribution functions $f_{\alpha}^{(eq)}$ in three-dimensional cases.

Once the DVS is determined, the velocity discretization of the distribution function $f_\alpha$ in Eq.~\eqref{BGK_a} can be performed. 
Specifically, multiplying both sides of Eq.~\eqref{BGK_a} by $w_k \omega^{-1}(\hat{\boldsymbol{\xi}}_{k})$ and introducing the discrete distribution function
\begin{equation}
    f_{\alpha,k} = w_k \omega^{-1}(\hat{\boldsymbol{\xi}}_{k})\, f_\alpha(\boldsymbol{\xi}_k),
\end{equation}
we obtain the discrete kinetic equations,
\begin{equation}
    \partial_t f_{\alpha,k}
+ {\boldsymbol{\xi}}_k \cdot \nabla f_{\alpha,k}
= -\frac{1}{\tau_f}\,\bigl[f_{\alpha,k} - f_{\alpha,k}^{(eq)}\bigr],
\label{new_kinetic1}
\end{equation}
where 
\begin{equation}
f_{\alpha,k}^{(eq)}=
\begin{cases}
 \rho\, \omega_k
   \Bigl\{
      1
      + \hat{\boldsymbol{\xi}}_k\!\cdot\!\hat{\boldsymbol{u}}
      + \tfrac{1}{2}
        \bigl[
            (\hat{\boldsymbol{\xi}}_k\!\cdot\!\hat{\boldsymbol{u}})^{2}
            - \hat{u}^{2}
            + (\theta - 1)(\hat{\xi}_k^{2} - D)
        \bigr]
   \Bigr\}, \qquad \alpha = 0,
\\[10pt]
 \rho \, \theta \, \omega_k
   \Bigl[
     (\hat{\boldsymbol{\xi}}_k\!\cdot\!\hat{\boldsymbol{u}}){\xi}_\alpha
      + {\xi}_\alpha - {u}_\alpha
   \Bigr]
   + {u}_\alpha \, f_{0,k}^{(eq)}, \qquad \alpha = 1,2,...,D.
\end{cases}
\end{equation}
% \begin{align}
% f_{0,k}^{(eq)}
% &
%  = \rho\, \omega_k
%    \Bigl\{
%       1
%       + \hat{\boldsymbol{\xi}}_k\!\cdot\!\hat{\boldsymbol{u}}
%       + \tfrac{1}{2}
%         \bigl[
%             (\hat{\boldsymbol{\xi}}_k\!\cdot\!\hat{\boldsymbol{u}})^{2}
%             - \hat{u}^{2}
%             + (\theta - 1)(\hat{\xi}_k^{2} - D)
%         \bigr]
%    \Bigr\},
% \\[10pt]
% \boldsymbol{f}_{k}^{(eq)}
% &
%  = \rho \, \theta \, \omega_k
%    \Bigl[
%      (\hat{\boldsymbol{\xi}}_k\!\cdot\!\hat{\boldsymbol{u}})\boldsymbol{\xi}
%       + \boldsymbol{\xi} - \boldsymbol{u}
%    \Bigr]
%    + \boldsymbol{u} \, f_{0,k}^{(eq)}.
% \end{align}
Accordingly, the macroscopic variables can be determined as,
\begin{equation}
\rho = \sum_{k} f_{0,k}, 
\qquad
\rho \boldsymbol{u} = \sum_{k} \boldsymbol{f}_{k},
\qquad
\boldsymbol{P} = \sum_{k} \boldsymbol{c}_{k}\, \boldsymbol{f}_{k},
\end{equation}
where $\boldsymbol{c}_{k} ={\boldsymbol{\xi}}_{k} - \boldsymbol{u}$.

Now we turn to Eq.~\eqref{BGK_b}. First, it is noted that the term $R$ in Eq.~\eqref{BGK_b} depends on the distribution function $f_\alpha$ and should be treated carefully. Here we adopt the approach proposed in Ref.~\cite{GUO2023Compressible},  in which $R$ is also represented by its Hermite expansion up to first order,
\begin{equation}
R \approx \omega(\hat{\boldsymbol{\xi}})\,(\boldsymbol{P} - \rho R T \boldsymbol{I}) : (\hat{\boldsymbol{\xi}} \hat{\boldsymbol{u}}),
\end{equation}
where $\boldsymbol{I}$ is the second-order unity tensor.

% It can be observed that the term $R$ in Eq.~\eqref{BGK_b} still explicitly involves the previously defined distribution function $f_\alpha$, 
% which introduces an inconsistency in the discretization procedure. 
% To overcome this issue, the approach proposed in Ref.~\cite{GUO2023Compressible} can be adopted, in which $R$ is treated through a Hermite expansion. 
On the other hand, since all distribution functions in the present work are only required to satisfy the moment conditions up to fourth-order accuracy, 
the same DVS can be employed for the discretization of $f_{D+1}$, and the final discrete velocity version of Eq.~\eqref{BGK_b} can be expressed as
\begin{equation}
    \partial_t f_{D+1,k}
    + {\boldsymbol{\xi}}_k \cdot \nabla f_{D+1,k}
    = -\frac{1}{\tau_e}\,\bigl[f_{D+1,k} - f_{D+1,k}^{(eq)}\bigr] 
      +\frac{1}{\tau_{ef}}R_k,
\label{new_kinetic2}
\end{equation}
where 
% $f_{D+1,k} = w_k \omega^{-1}(\hat{\boldsymbol{\xi}}_{k})f_{D+1}(\boldsymbol{\xi}_k)$, $R_k = \sum_\alpha(f_{\alpha,k}-f_{\alpha,k}^{(eq)})u_\alpha $, and 
\begin{align}
    R_k &= w_k\,(\boldsymbol{P} - p \boldsymbol{I}) : (\hat{\boldsymbol{\xi}}_k \hat{\boldsymbol{u}}), \label{R_k} \\
    f_{D+1,k} &= w_k \omega^{-1}(\hat{\boldsymbol{\xi}}_{k})f_{D+1}(\boldsymbol{\xi}_k), \\
    % R_k &= (\boldsymbol{f}_{k}-\boldsymbol{f}_{k}^{(eq)}) \cdot \boldsymbol{u}, \label{R_k} \\
    f_{D+1,k}^{(eq)}
        &= \omega_k\,
        p\Bigl[
        (\hat{\boldsymbol{\xi}}_k\!\cdot\!\hat{\boldsymbol{u}})
        + (\hat{\boldsymbol{\xi}}_k\!\cdot\!\hat{\boldsymbol{u}})^{2}
        - \hat{u}^{2}
        + \tfrac{1}{2}\theta(\hat{\xi}_k^{2} - D)
     \Bigr]
     + E\, f_{0,k}^{(eq)}.
\end{align}
Accordingly, the energy and heat flux are given by
\begin{equation}
\rho E = \sum_{k} f_{D+1,k}, 
\qquad
\boldsymbol{q} = \sum_{k} \boldsymbol{c}_{k}\, (f_{D+1,k} - \boldsymbol{f}_{k} \cdot \boldsymbol{u}) = \sum_{k} \boldsymbol{c}_{k}\, f_{D+1,k} - \boldsymbol{P} \cdot \boldsymbol{u}.
\end{equation}

\section{DUGKS FOR THE VECTORIAL KINETIC MODEL}\label{sec:3}
\subsection{Numerical method}
Equations \eqref{new_kinetic1} and \eqref{new_kinetic2} have the same structure, which can be written as
\begin{equation}
    \partial_t \phi_i + \boldsymbol{\xi}_i \cdot \nabla\phi_i = Q_i^{\phi},
    \label{finaleq}
\end{equation}
where $\phi = f_\alpha$, with $ \alpha=0,1,...,D+1$ and 
% the corresponding collision term $Q_i^\phi$ is defined as
\begin{equation}
   Q_i^{\phi} =
\begin{cases}
-\frac{1}{\tau_f} \left[ f_{\alpha ,i} - f_{\alpha,i}^{{(eq)}} \right], 
& \alpha = 0,1,...,D,  \\[10pt]
-\frac{1}{\tau_e} \left[ f_{\alpha,i} - f_{\alpha,i}^{{(eq)}} \right] + \frac{1}{\tau_{ef}} R_i, 
& \alpha = {D+1}.
\end{cases}
\end{equation}
% \begin{equation}
%    Q_i^{\phi} =
% \begin{cases}
% -\frac{1}{\tau_f} \left[ f_{\alpha ,i} - f_{\alpha,i}^{{(eq)}} \right], 
% & \phi = f_\alpha,\quad (\alpha = 0,1,...,D)  \\[10pt]
% -\frac{1}{\tau_e} \left[ f_{D+1,i} - f_{D+1,i}^{{(eq)}} \right] + \frac{1}{\tau_{ef}} R_i, & \phi = f_{D+1},
% \end{cases}
% \end{equation}
Here, the coupling term $R$ is discretized following Eq.~\eqref{R_k}, and the collision operator $Q_i^\phi$ satisfies the zeroth-order conservation 
% for all $\phi = f_\alpha$, i.e.,
\begin{equation}
    \sum_i Q_i^\phi = 0.
    \label{collision_invariant}
\end{equation}

In the following, the DUGKS is employed to solve Eq.~\eqref{finaleq}. 
We first perform the spatial discretization using the finite-volume framework, in which the computational domain is partitioned into a set of control volumes $V_j$ centered at $\boldsymbol{x}_j$. 
Accordingly, the cell-averaged distribution function over each control volume is introduced to represent the macroscopic information within the cell,
\begin{equation}
\phi_{i,j}(t) = \frac{1}{|V_j|} \int_{V_j} \phi_i(\boldsymbol{x}, t) \, d \boldsymbol{x} ,
\end{equation}
where $|V_j|$ denotes the volume of the control volume $V_j$. 
Integrating Eq.~\eqref{finaleq} over the control volume $V_j$ and over the time interval from $t_n$ to $t_{n+1}$ yields
\begin{equation}
\phi_{i,j}^{n+1} - \phi_{i,j}^n 
+ \frac{\Delta t}{|V_j|} \mathcal{F}_{i,j}^{\phi, n+1/2} 
= \frac{\Delta t}{2} \left[ Q_{i,j}^{\phi, n+1} + Q_{i,j}^{\phi, n} \right],
\label{7}
\end{equation}
where the time integral of the collision term is approximated by the trapezoidal rule, while the flux term is evaluated using the midpoint rule. 
It can be observed that the resulting update equation for $\phi_{i,j}$ is implicit in time. 
To remove this implicitness, two auxiliary distribution functions are introduced through the following transformation:
\begin{equation}
    \tilde{\phi}_i = \phi_i - \frac{\Delta t}{2} Q_i^{\phi}, \qquad 
    \tilde{\phi}_i^+ = \phi_i + \frac{\Delta t}{2} Q_i^{\phi}.
    \label{8}
\end{equation}
With this transformation, Eq.~\eqref{7} can be implemented explicitly,
\begin{equation}
    \tilde{\phi}_{i,j}^{n+1} = \tilde{\phi}_{i,j}^{+,n} - \frac{\Delta t}{|V_j|} \mathcal{F}_{i,j}^{\phi, n+1/2}.
    \label{11}
\end{equation}

According to the collision invariants described in Eq.~\eqref{collision_invariant}, mass, momentum, and energy are preserved during particle collisions. 
As a result, the macroscopic variables can be directly obtained from the auxiliary distribution function $\tilde{\phi}_i$. 
In particular, the macroscopic variables can be calculated as
\begin{equation}
\rho = \sum_i \tilde{f}_{0,i},
\qquad
\rho \boldsymbol{u} = \sum_i \tilde{\boldsymbol{f}}_i ,
\qquad
\rho E = \sum_i \tilde{f}_{D+1,i}.
\end{equation}
Consequently, the auxiliary distribution $\tilde{\phi}_i$ is macroscopically equivalent to the original distribution function, and can therefore be employed in practical simulations.

The remaining and most crucial task is to calculate the interface flux $\mathcal{F}_{i,j}^{\phi,\,n+1/2}$ according to the following relation:
\begin{equation}
    \mathcal{F}_{i,j}^{\phi,n+1/2} = \int_{\partial V_j} ({\boldsymbol{\xi}}_i \cdot \boldsymbol{n}) \phi_i(\boldsymbol{x}, t_{n+1/2}) \, dS
    = \sum_{k}
   \left( {\boldsymbol{\xi}}_i \cdot \boldsymbol{n}_{b} \right)
   \phi_i^{\,n+1/2}(\boldsymbol{x}_{b})\, S_{b}.
   \label{flux}
\end{equation}
% From the above expression, it is clear that the evaluation of the flux requires 
Therefore, the key is to evaluate the distribution function $\phi_i$ at the cell interface at the half time $t_{n+1/2}$. 
In DUGKS, this is achieved by integrating Eq.~\eqref{finaleq} along the characteristic line over a half time step $h = \Delta t / 2$,
% During this integration, the collision term is approximated by the trapezoidal rule, which leads to
\begin{equation}
\phi_i(\boldsymbol{x}_b, t_n + h) - \phi_i(\boldsymbol{x}_b - \boldsymbol{\xi}_i h,  t_n) = \frac{h}{2} [Q_i^\phi(\boldsymbol{x}_b, t_n + h) + Q_i^\phi(\boldsymbol{x}_b - \boldsymbol{\xi}_i h, t_n)],
\label{13}
\end{equation}
where the time integration of the collision term is approximated by the trapezoidal rule. By introducing the following two auxiliary distribution functions,
% In analogy with the previous treatment, the distribution function is transformed once again to remove the implicit terms in the governing equation:
\begin{equation}
    \bar{\phi_i} = \phi_i - \frac{h}{2}Q_i^\phi, \qquad
    \bar{\phi}_i^+ = \phi_i + \frac{h}{2}Q_i^\phi,
    \label{eq:bianhuan2}
\end{equation}
and then the implicitness can be removed.
Consequently, Eq.~\eqref{13} is rewritten as
\begin{equation}
    \bar{\phi_i}(\boldsymbol{x}_b, t_n + h) = \bar{\phi}_i^+(\boldsymbol{x}_b - \boldsymbol{\xi}_i h, t_n).
    \label{15}
\end{equation}

\begin{figure}[htbp]
    \centering
    \includegraphics[width=0.65\textwidth]{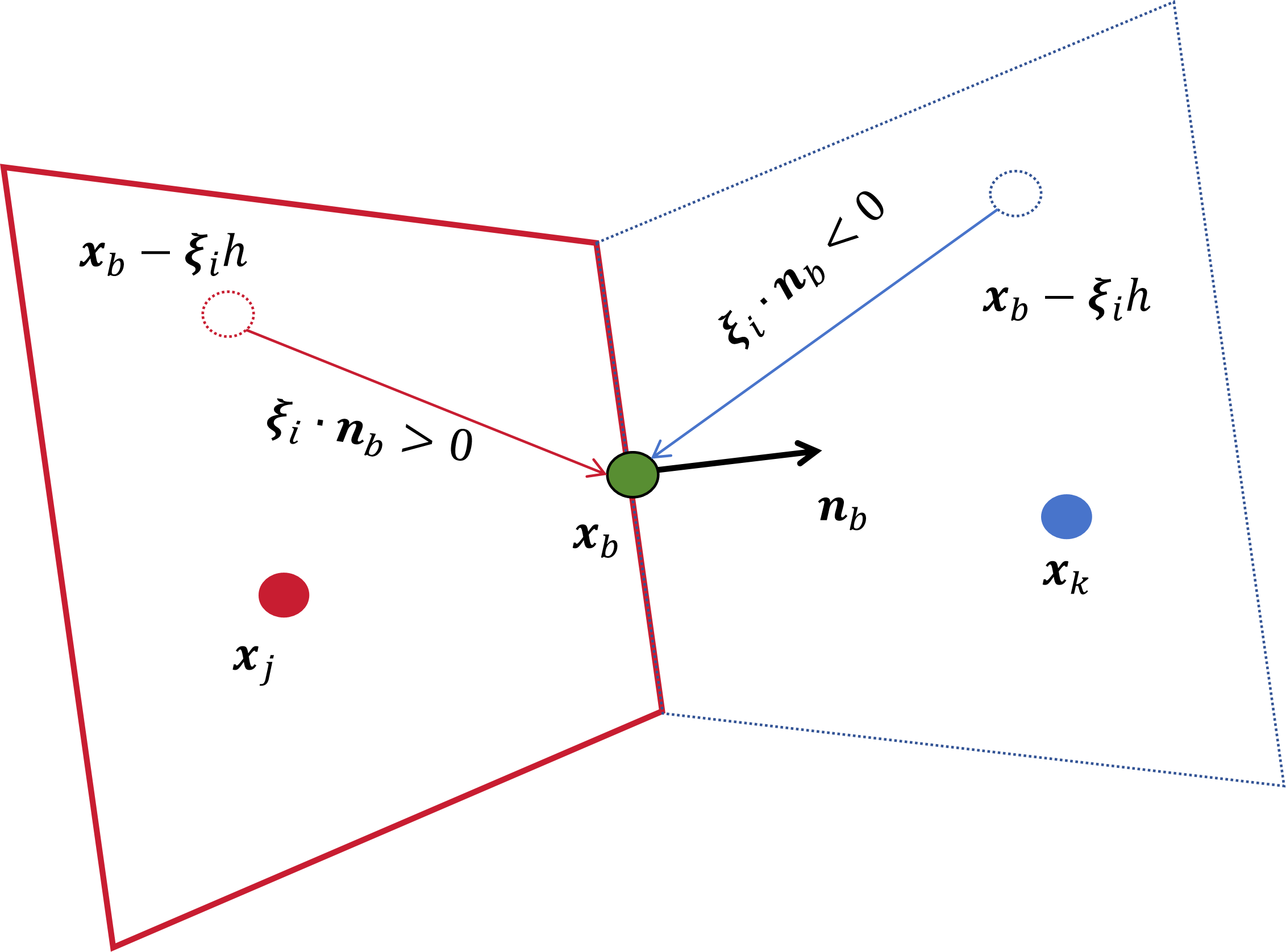}
    \caption{Schematic illustration of the interface flux evaluation in the V-DUGKS for two adjacent control volumes $V_j$ and $V_k$ with centers at $\boldsymbol{x}_j$ and $\boldsymbol{x}_k$, respectively.
Here, $\boldsymbol{x}_b$ is the interface point, and $\boldsymbol{n}_b$ denotes the outward unit normal of $V_j$ at $\boldsymbol{x}_b$.
% The shifted points $\boldsymbol{x}_b-\boldsymbol{\xi}_i h$ along the discrete velocity direction are used for the characteristic-based reconstruction.
}
    \label{fig:DUGKS}
\end{figure}

The next key step is the evaluation of the right-hand-side terms in Eq.~\eqref{15}. 
To this end, the auxiliary distribution function $\bar{\phi}_i^{+}(\boldsymbol{x}_b - \boldsymbol{\xi}_i h, t_n)$ is reconstructed as follows
\begin{equation}
\bar{\phi}_{i}^{+}(\boldsymbol{x}_b - \boldsymbol{\xi}_i h, t_n)
=
\begin{cases}
\bar{\phi}_{i}^{+}(\boldsymbol{x}_j, t_n)
+ (\boldsymbol{x}_{b} - {\boldsymbol{\xi}}_{i} h - \boldsymbol{x}_j)\,
  \sigma_{i,j},
& {\boldsymbol{\xi}}_i \cdot \boldsymbol{n}_{b} > 0,
\\[6pt]
\bar{\phi}_{i}^{+}(\boldsymbol{x}_k, t_n)
+ (\boldsymbol{x}_{b} - {\boldsymbol{\xi}}_{i} h - \boldsymbol{x}_k)\,
  \sigma_{i,k},
& {\boldsymbol{\xi}}_i \cdot \boldsymbol{n}_{b} < 0,
\end{cases}
\end{equation}
where $\boldsymbol{\sigma}_{i,\gamma}$ denotes the slope of $\bar{\phi}_i^{+}$ within the control volume $V_\gamma$ ($\gamma = j$ or $k$). 
As an example, in Fig. \ref{fig:DUGKS} a 2D case is shown.
In the present study, this slope is computed using the van Leer limiter~\cite{VANLEER1977} to enhance numerical stability. 
For the one-dimensional case, the slope of $\bar{\phi}^{+}$ in cell $V_j$ is given by
\begin{equation}
\sigma_j(\bar{\phi}^{+})
=
\frac{\operatorname{sgn}(s_1) + \operatorname{sgn}(s_2)}{|s_1| + |s_2|}
\, |s_1|\,|s_2| ,
\end{equation}
where
\begin{equation}
s_1 = \frac{\bar{\phi}_{i,j}^{+} - \bar{\phi}_{i,j-1}^{+}}{x_j - x_{j-1}},
\qquad
s_2 = \frac{\bar{\phi}_{i,j+1}^{+} - \bar{\phi}_{i,j}^{+}}{x_{j+1} - x_j},
\end{equation}
with ${x}_j$ being the center of cell $V_j$.

Once the interface distribution function $\bar{\phi}_i(\boldsymbol{x}_b, t_n + h)$ is obtained, the conservative macroscopic variables at the half time level can be evaluated according to
\begin{equation}
\rho^{\,n+1/2}
= \sum_{i} \bar{f}_{0,i}^{\,n+1/2},
\qquad
(\rho \boldsymbol{u})^{\,n+1/2}
= \sum_{i} \bar{\boldsymbol{f}}_{i}^{\,n+1/2},
\qquad
(\rho E)^{\,n+1/2}
= \sum_{i} \bar{f}_{D+1,i}^{\,n+1/2},
\end{equation}
the corresponding pressure tensor is then given by
\begin{equation}
\boldsymbol{P}^{\,n+1/2}
=
\frac{4\tau_{f}}{4\tau_{f} + \Delta t}
\sum_{i}
\boldsymbol{c}_{i} \, \bar{\boldsymbol{f}}_{i}^{\,n+1/2} + 
\frac{\Delta t}{4\tau_{f} + \Delta t} p^{n+1/2} \boldsymbol{I}.
\end{equation}
Then with the help of Eq.~\eqref{eq:bianhuan2}, the required original distribution function $\phi_i(\boldsymbol{x}_b, t_n + h)$ can be readily obtained,
\begin{equation}
    \phi_i(\boldsymbol{x}_b, t_n + h) = \frac{2\tau_\phi}{2\tau_\phi + h} \bar{\phi_i}(\boldsymbol{x}_b, t_n + h) + \frac{h}{2\tau_\phi + h} \left[\phi_i^{(eq)}(\boldsymbol{x}_b, t_n + h) + S_i^\phi(\boldsymbol{x}_b, t_n + h) \right],
    \label{flux}
\end{equation}
with
\begin{equation}
    S_i^\phi = 
    \begin{cases}
        0, \quad \phi=f_\alpha,\quad \alpha = 0,1,...,D,\\
        (1-\frac{1}{\mathrm{Pr}})R_i, \quad \phi = f_{D+1},
    \end{cases}
\end{equation}
and
\begin{equation}
    \tau_\phi = 
    \begin{cases}
        \tau_f, \quad \phi=f_\alpha,\quad \alpha = 0,1,...,D,\\
        \tau_e, \quad \phi = f_{D+1}.
    \end{cases}
\end{equation}

At this stage, the interface flux $\mathcal{F}_{i,j}^{\phi,\,n+1/2}$ can be evaluated according to Eq.~\eqref{flux}. 
The distribution function $\tilde{\phi}_i$ is then updated using Eq.~\eqref{11}. 
During this update procedure, the following relations are additionally employed:
\begin{equation}
\bar{\phi}_{i}^{+}
=
\frac{4\tau_\phi - \Delta t}{4\tau_\phi + 2\Delta t}\,
\tilde{\phi}_{i}
+
\frac{3\Delta t}{4\tau_\phi + 2\Delta t}\,
(\phi_{i}^{(eq)} + S_i^\phi),
\qquad
\tilde{\phi}_{i}^{+}
=
\frac{4}{3}\,\bar{\phi}_{i}^{+}
-
\frac{1}{3}\,\tilde{\phi}_{i},
\end{equation}
and the pressure tensor and heat flux are evaluated as
\begin{align}
\boldsymbol{P}
&=
\frac{2\tau_{f}}{2\tau_{f} + \Delta t}
\sum_{i} \boldsymbol{c}_{i}\, \tilde{\boldsymbol{f}}_{i}
\;+\;
\frac{\Delta t}{2\tau_{f} + \Delta t}
\, p\, \boldsymbol{I},\\
\boldsymbol{q}
&= \sum_{i} \boldsymbol{c}_{i}\, \tilde{f}_{D+1,i}
- \boldsymbol{P} \cdot \boldsymbol{u}.
\end{align}

Finally, the time step $\Delta t$ is selected according to the Courant–Friedrichs–Lewy (CFL) stability criterion, which is given by
\begin{equation}
\Delta t
=
\frac{\beta \, \Delta x}
{|\boldsymbol{\xi}|_{\max} + |\boldsymbol{u}|_{\max} + c_{\max}} .
\label{CFL}
\end{equation}
Here, $\beta$ denotes the CFL number, the parameter $\Delta x$ is the spatial mesh size,
% satisfying $0 < \beta \le \beta_0$, where $\beta_0$ is a prescribed constant. 
$|\boldsymbol{\xi}|_{\max}$ represents the maximum discrete particle speed among the employed DVS in table \ref{tab:DVS}, 
$|\boldsymbol{u}|_{\max}$ is the maximum macroscopic flow velocity, and $c_{\max}$ denotes the maximum sound speed in the computational domain.

\subsection{Computational efficiency analysis of the V-DUGKS}
In addition to numerical accuracy and stability, computational efficiency is a critical consideration for kinetic schemes applied to continuum compressible flows.
From a methodological perspective, the computational efficiency of the proposed V-DUGKS can be understood by examining the required moment conditions, the associated Hermite expansion, and the resulting discrete velocity representation.

Firstly, it can be observed from Eq.~\eqref{moments} that, within the V-DUGKS framework, only second--order moment conditions are required to correctly recover the macroscopic conservative variables.
This is in contrast to the formulation in Ref.~\cite{GUO2023Compressible}, where third--order moment conditions are involved in the evaluation of the velocity field.
As a direct consequence, the equilibrium distribution functions~\eqref{hermite} in V-DUGKS are truncated at the second order in the Hermite expansion, and third--order Hermite polynomials are entirely avoided.
This simplification leads to more compact equilibrium expressions and reduces the computational effort associated with evaluating high--order velocity moments, thereby lowering the per--time--step computational complexity.

Moreover, owing to the second--order Hermite truncation, higher--order terms such as $\hat{\boldsymbol{\xi}}^3$ and $\hat{\boldsymbol{u}}^3$ do not appear in the equilibrium formulation.
Accordingly, coefficients involving high--order inverse powers of the reference temperature, e.g.\ $(R T_{{ref}})^{-3/2}$, are eliminated.
Since such terms may amplify numerical perturbations when $T_{{ref}}$ deviates from the local temperature, their removal effectively alleviates the sensitivity of the scheme to the choice of the reference temperature.
This feature not only simplifies the numerical implementation but also contributes to improved robustness and stability, particularly for high--Mach--number flows.

% In addition, the reduction in the required order of moment conditions relaxes the accuracy requirement on the Gauss--Hermite quadrature.
% Within the V-DUGKS framework, a discrete velocity set with fifth--order accuracy is sufficient to ensure the exact recovery of the retained moments.
% Lower--order discrete velocity sets are generally associated with smaller maximum discrete velocities $|\boldsymbol{\xi}|_{\max}$.
% According to the CFL constraint, a smaller $|\boldsymbol{\xi}|_{\max}$ permits a larger allowable time step.
% As a result, for a given physical simulation time, V-DUGKS can advance the solution with fewer time--marching steps, which is a key factor contributing to its overall computational efficiency.

In addition, the reduction in the required order of moment conditions leads to a lower accuracy requirement for the Gauss--Hermite quadrature.
In the present work, a DVS with fifth-order accuracy is sufficient.
As shown in Table~\ref{tab:DVS}, the maximum discrete velocity magnitude $|\boldsymbol{\xi}|_{\max}$ associated with lower-order DVS is significantly smaller than that of higher-order models.
According to the CFL condition in Eq.~\eqref{CFL}, a smaller $|\boldsymbol{\xi}|_{\max}$ allows for a larger time step $\Delta t$.
Consequently, for a given physical simulation time, the V-DUGKS requires fewer time-marching steps, resulting in reduced computational time and improved overall efficiency.

It is worth noting that, although the vectorial formulation of V-DUGKS may involve a larger total number of discrete velocity points compared with some conventional scalar formulations, this does not necessarily imply a higher overall computational cost.
The allowance of a larger time step and the reduced number of time steps required to reach a prescribed physical time can effectively compensate for the additional cost associated with the increased velocity degrees of freedom.
Therefore, from a theoretical standpoint, V-DUGKS is expected to achieve competitive, and potentially superior, computational efficiency compared with the conventional DUGKS.

\section{NUMERICAL TESTS}\label{sec:4}
In this section, a series of numerical tests are carried out to assess
the accuracy, stability, and computational efficiency of the proposed V-DUGKS
for continuum compressible flows. 
The test cases include the one–dimensional shock-tube problems, the Shu--Osher
problem, a two–dimensional Riemann problem, and the three–dimensional Taylor--Green vortex. These benchmarks are selected to examine the capability
of the proposed scheme in capturing strong discontinuities and preserving accuracy in smooth regions.
In addition, the computational efficiency of V-DUGKS is evaluated through
comparisons with the standard DUGKS under identical numerical settings.
Here, the standard DUGKS refers to the conventional scalar-distribution-function DUGKS for continuum compressible flows and is used as the baseline method in the following comparisons.
Unless otherwise specified, the van Leer limiter is employed in the
reconstruction of interface fluxes, and the CFL number is fixed throughout the
simulations. 

\subsection{Sod shock tube}
To test the numerical stability of the proposed scheme in the presence of strong discontinuities, a set of one--dimensional shock tube problems is considered. As a first benchmark, we investigate the classical Sod shock tube problem~\cite{SOD19781}, the computational domain is defined as $x \in [-0.5,\,0.5]$ and the flow is initially separated by a diaphragm located at $x=0$, 
% The macroscopic variables are prescribed as
\begin{equation}
(\rho,u,p)=
\begin{cases}
(\rho_L,u_L,p_L)=(1.0,\,0.0,\,1.0), & x\le 0,\\[4pt]
(\rho_R,u_R,p_R)=(0.125,\,0.0,\,0.1), & x>0.
\end{cases}
\label{eq:sod_init}
\end{equation}
% The gas is modeled as a hard--sphere gas, for which 
The dynamic viscosity depends on the temperature as $\mu = \mu_0 \left( {T}/{T_0} \right)^{0.5}$, where $\mu_0$ denotes the reference viscosity at the reference temperature $T_0$. To remain within the continuum compressible regime, the reference viscosity is set to $\mu_0 = 5\times10^{-5}$. The reference density, pressure, and temperature are taken from the upstream state, i.e.,
\begin{equation}
\rho_0=\rho_L,\qquad p_0=p_L,\qquad T_0=T_L.
\end{equation}

For the purpose of assessing the stability, the numerical reference temperature used in the Hermite expansion is set to $T_{{ref}}=0.8$. The ratio of specific heats is fixed at $\gamma=1.4$, and the Prandtl number is set to $\mathrm{Pr}=2/3$. A uniform grid with $N=600$ cells is employed for the spatial discretization. 
For the velocity space, the standard DUGKS adopts the D1V4--A7 set for each distribution function (i.e., $4+4$ discrete velocities), whereas the proposed V-DUGKS adopts the D1V3--A5 set consisting of $3\times3$ discrete velocity points. 
% The DVSs specified above are used throughout all subsequent one--dimensional simulations unless otherwise noted.
 The CFL number is fixed at $0.95$, and the simulation is advanced up to $t=0.15$, where comparisons are made with the exact Riemann solution.

\begin{figure}[h]
    \centering
    % -------- first row --------
    \begin{subfigure}{0.45\textwidth}
        \centering
        \includegraphics[width=\textwidth]{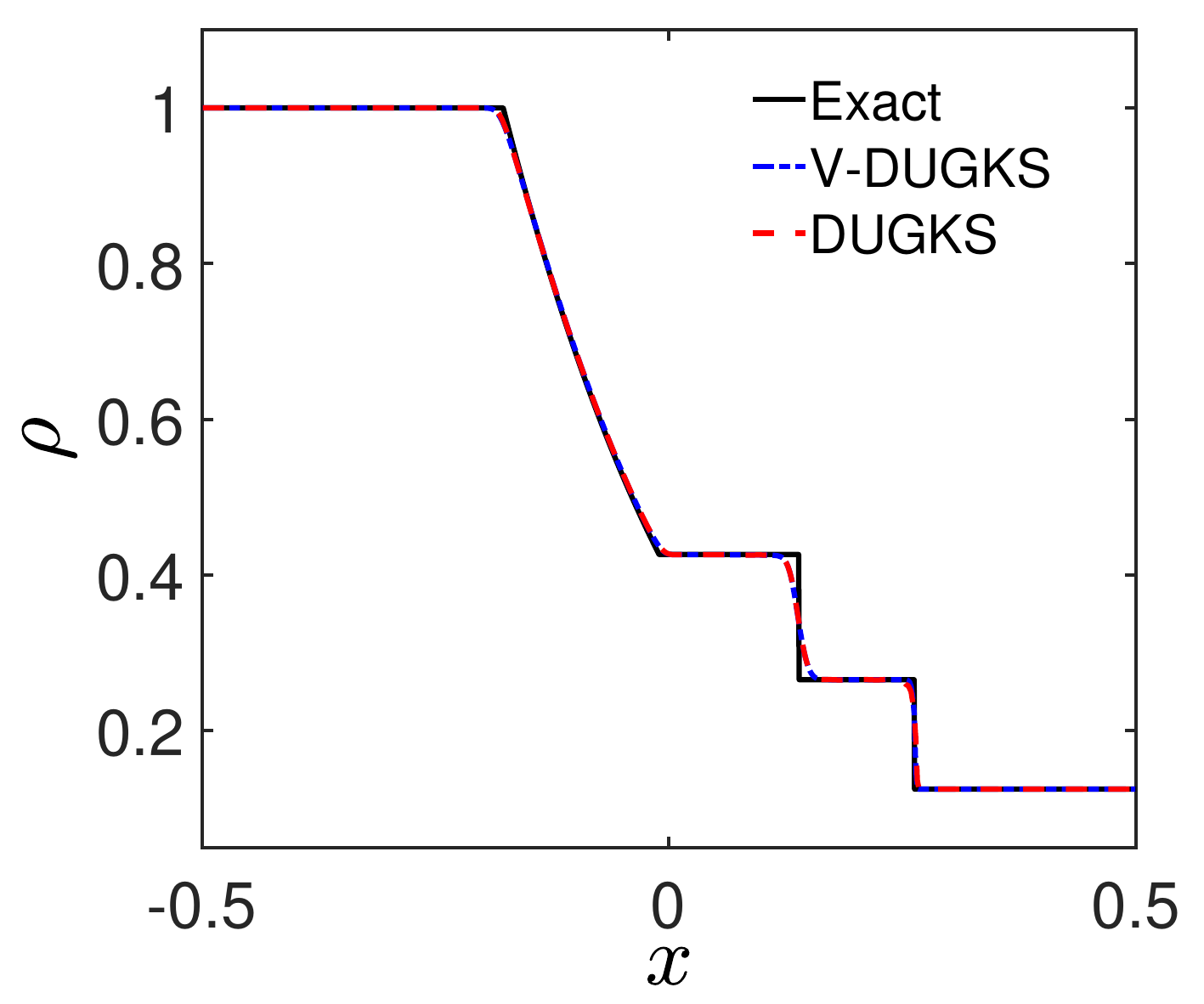}
        \caption{Density}
    \end{subfigure}
    \hfill
    \begin{subfigure}{0.45\textwidth}
        \centering
        \includegraphics[width=\textwidth]{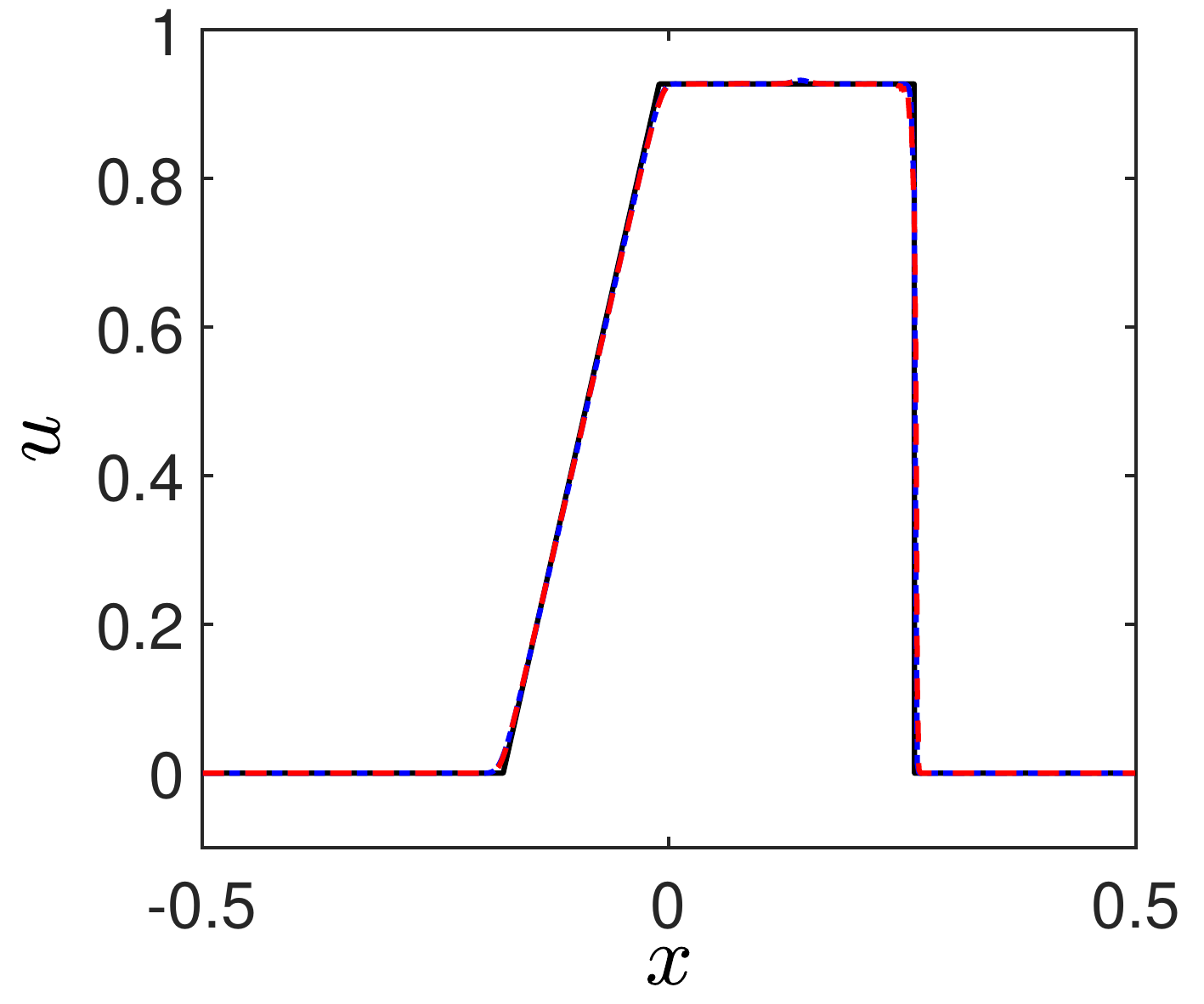}
        \caption{Velocity}
    \end{subfigure}

    % \vspace{0.8em} % 行间距，可微调

    % -------- second row --------
    \begin{subfigure}{0.62\textwidth}
        \centering
        \includegraphics[width=\textwidth]{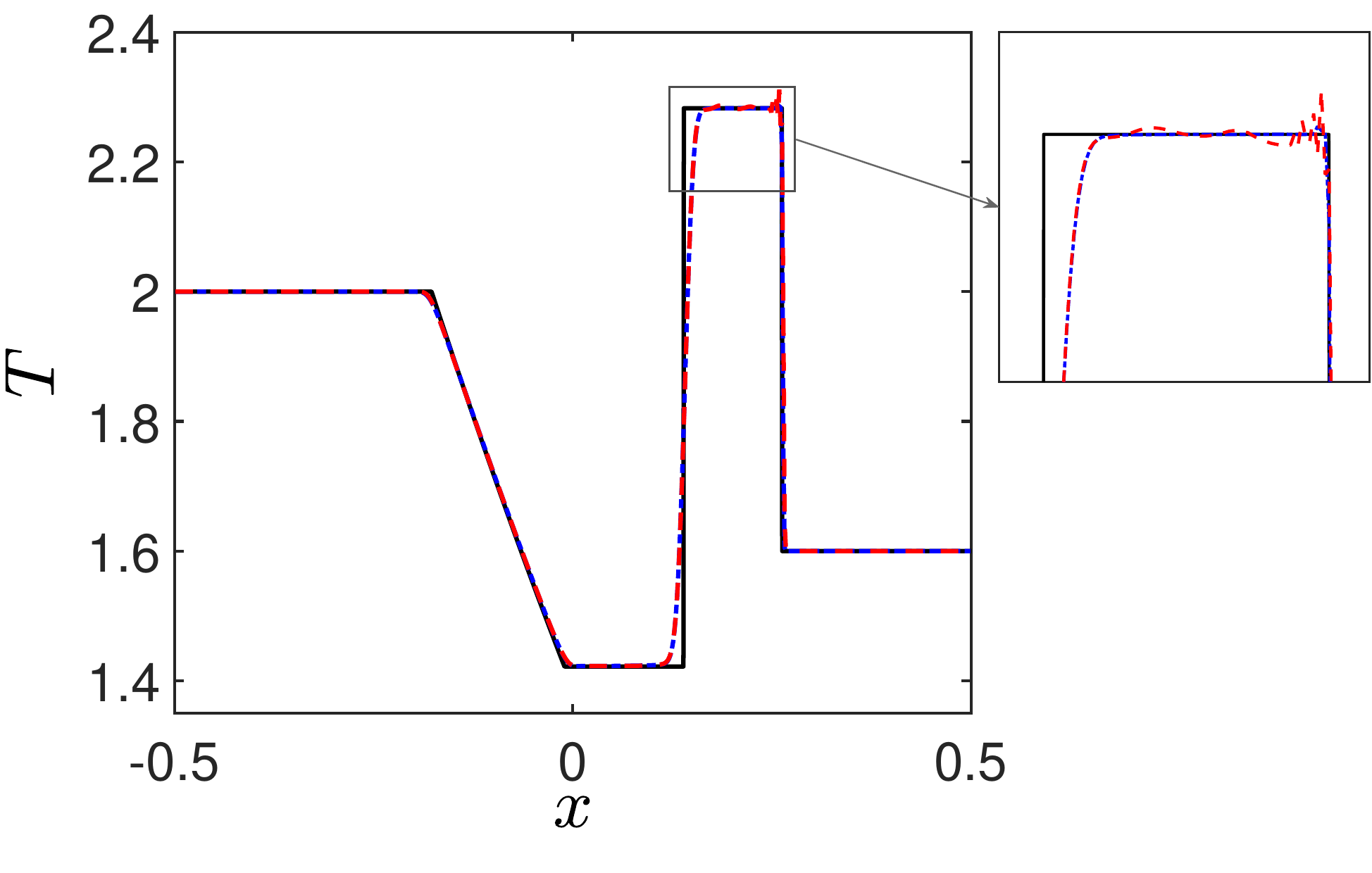}
        \caption{Temperature}
    \end{subfigure}
    \caption{Numerical results predicted by the standard DUGKS and the present V-DUGKS of the Sod shock-tube problem with 600 cells at $t=0.15$.}
    \label{fig:sod_profiles}
\end{figure}

The density, velocity, and temperature profiles obtained from the standard DUGKS and V-DUGKS are shown in Fig.~\ref{fig:sod_profiles}. It can be seen that both schemes accurately capture the shock wave, contact discontinuity, and rarefaction fan, and show good agreement with the exact solution. Notably, the temperature profile, which is the most sensitive quantity in this test, exhibits improved robustness when computed by the V-DUGKS, as evidenced by the reduced spurious oscillations near the contact discontinuity.
Further numerical experiments indicate that the V-DUGKS remains stable when the numerical reference temperature $T_{{ref}}$ is further decreased to $0.7$, whereas the standard DUGKS fails to converge under the same conditions. This demonstrates that the proposed V-DUGKS is less sensitive than the standard DUGKS to the choice of $T_{{ref}}$. The enhanced stability can be attributed to the modified equilibrium distribution used in the V-DUGKS, in which the third-order Hermite terms are discarded, thereby eliminating the strong dependence on the cubic power of the thermal velocity scale $\sqrt{R T_{{ref}}}$. As a result, improved numerical stability is achieved without sacrificing accuracy.

\subsection{Lax shock tube}
As a more stringent test for numerical stability, the Lax shock tube problem~\cite{Lax1954WeakSO} is considered next.
The initial condition is prescribed as
\begin{equation}
(\rho,u,p)=
\begin{cases}
(\rho_L,u_L,p_L) = (0.445,\;0.698,\;3.528), & x \le 0,\\
(\rho_R,u_R,p_R) = (0.5,\;0.0,\;0.571), & x > 0,
\end{cases}
\end{equation}
and the computational domain is $x\in[-0.5,\,0.5]$.
% The gas is modeled as hard--sphere molecules, and the dynamic viscosity is evaluated by $\mu=\mu_0 (T/T_0)^{0.5}$ with 
The gas is the same as in the Sod problem, and we set $\mu_0=1.5\times10^{-4}$, where the reference temperature $T_0$ corresponds to the upstream state. In the Hermite expansion, the numerical reference temperature is set to $T_{{ref}}=4.7$ to ensure adequate convergence. The Prandtl number is fixed at $\mathrm{Pr}=2/3$, and a uniform mesh with $500$ grid points is employed. All simulations are carried out with a CFL number of $0.95$, and the solution is advanced to $t=0.12$.
\begin{figure}[h]
    \centering
    % -------- First row --------
    \begin{subfigure}{0.45\textwidth}
        \centering
        \includegraphics[width=\textwidth]{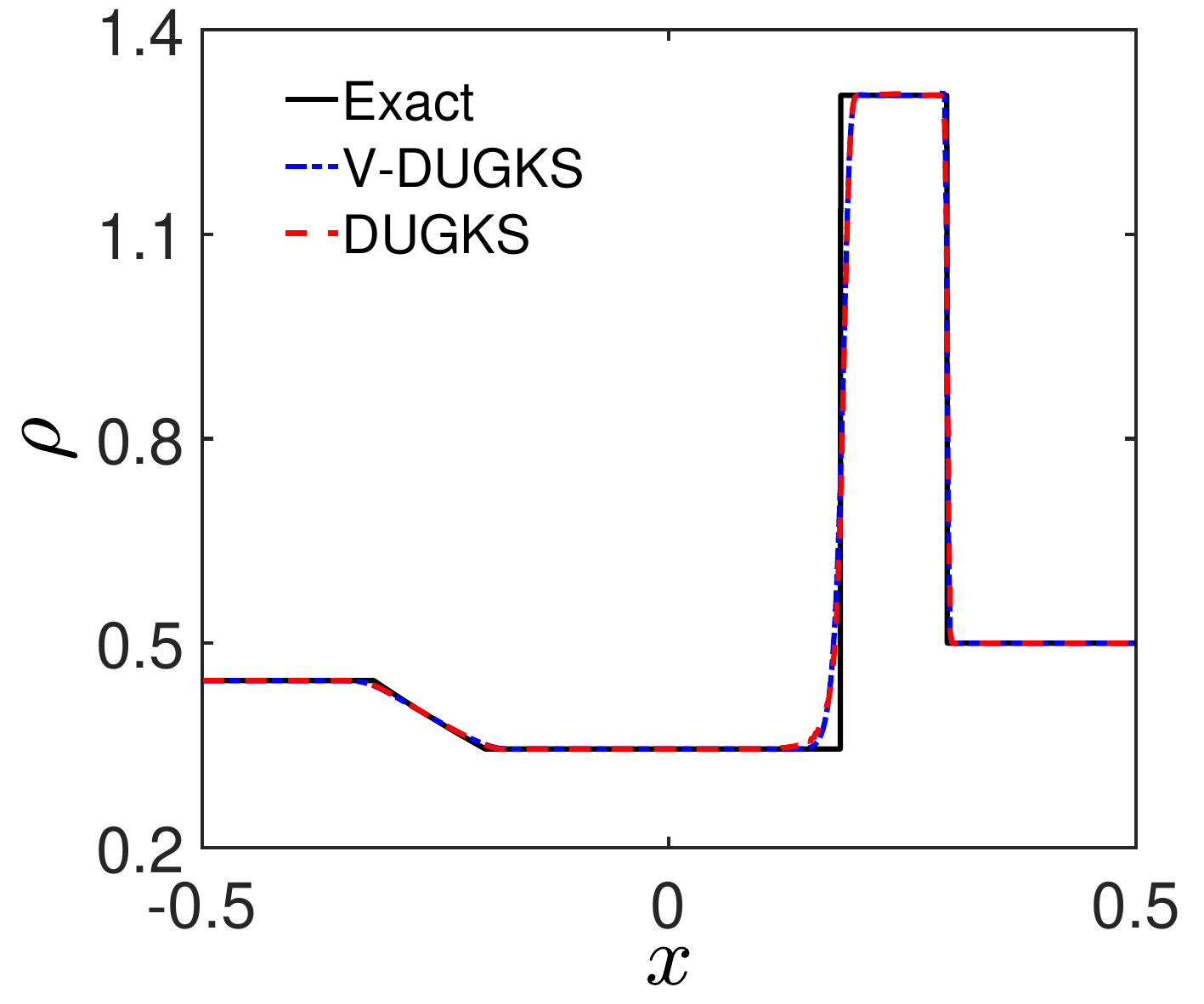}
        \caption{Density}
    \end{subfigure}
    \hfill
    \begin{subfigure}{0.45\textwidth}
        \centering
        \includegraphics[width=\textwidth]{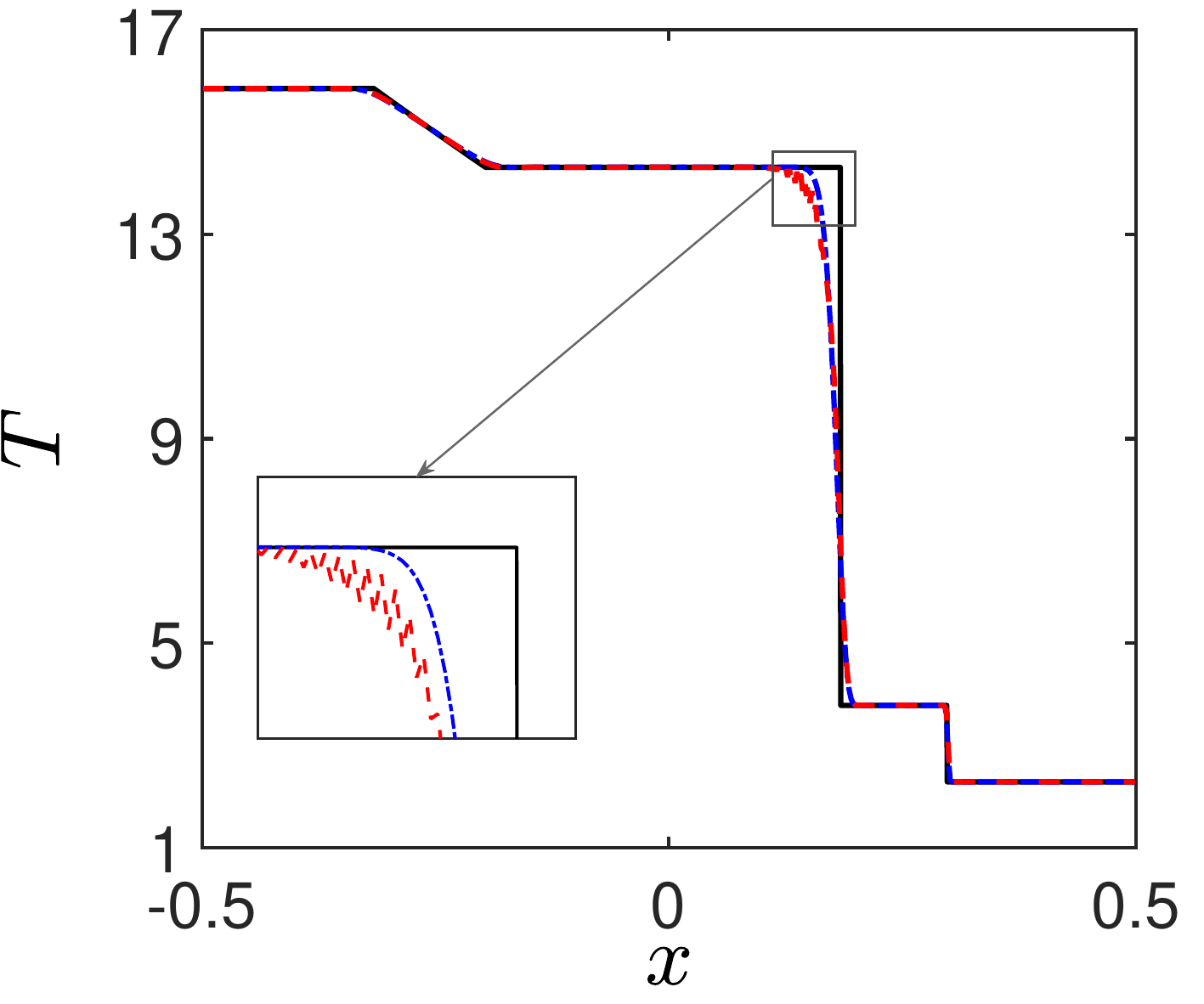}
        \caption{Temperature}
    \end{subfigure}

    % \vspace{-0.8em}
    % -------- Second row --------
    \begin{subfigure}{0.62\textwidth}
        \centering
        \includegraphics[width=\textwidth]{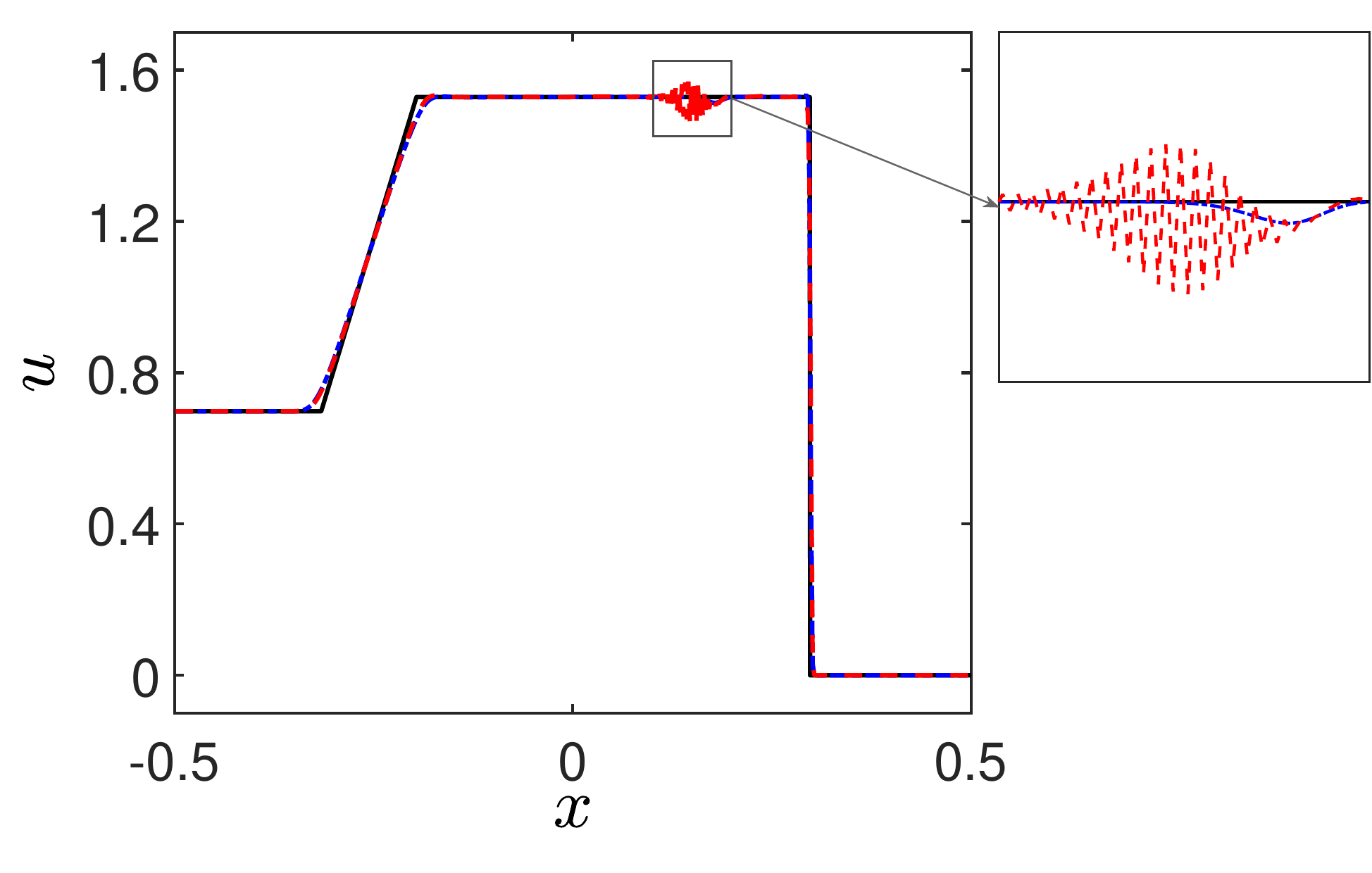}
        \caption{Velocity}
    \end{subfigure}
    
    \caption{Numerical results predicted by the standard DUGKS and the present V-DUGKS of the Lax shock-tube problem with 500 cells at $t=0.12$.}
    \label{fig:lax_profiles}
\end{figure}

Figure~\ref{fig:lax_profiles} compares the density, temperature, and velocity profiles predicted by the standard DUGKS and the proposed V-DUGKS for this problem.
It can be seen that both schemes are able to capture the main flow structures, including the shock wave,
contact discontinuity, and rarefaction wave, and show good agreement with the exact Riemann solution. Nevertheless, noticeable differences arise in the vicinity of strong discontinuities. In particular, the standard DUGKS exhibits visible oscillations in the velocity and temperature profiles near the shock front, as highlighted by the enlarged view, whereas the V-DUGKS maintains a smooth and stable solution under the same numerical conditions.

These results clearly demonstrate that the proposed V-DUGKS possesses enhanced numerical stability for strong compressible flows. 
It is also noted that as a smaller numerical reference temperature $T_{{ref}}$ is employed in the Hermite expansion, e.g., $T_{ref}=3.7$, the contribution of high--order terms in the equilibrium distribution becomes more pronounced, which can adversely affect the stability of the standard DUGKS, especially when combined with the trapezoidal time integration of the collision term. 
In contrast, the V-DUGKS effectively mitigates the
influence of these high--order contributions through its modified formulation, thereby exhibiting improved robustness with respect to the choice of $T_{{ref}}$ and maintaining stable solutions for this problem.

\subsection{Shu-Osher problem}
Finally, the Shu--Osher problem~\cite{SHUOsher1988439} is adopted to examine the stability of the proposed V-DUGKS.
This problem involves the interaction between a shock wave and a sinusoidal density perturbation, and is widely used for stability assessment in compressible flow simulations.
The computation is performed in a one-dimensional domain $x \in [-5,5]$, with the initial condition given by
\begin{equation}
(\rho,u,p)=
\begin{cases}
(\rho_L,u_L,p_L) = (3.857143,\; 2.629369,\; 10.3333333333), & x<-4, \\[4pt]
(\rho_R,u_R,p_R) = \bigl(1+0.2\sin(5x),\; 0,\; 1\bigr), & x\ge -4.
\end{cases}
\end{equation}
The ratio of specific heats is taken as $\gamma=1.4$.
The simulations are performed on a uniform mesh consisting of $N=800$ grid points, with zero-gradient boundary conditions imposed at both ends of the domain.
The reference viscosity and numerical reference temperature are set to $\mu_{0}=10^{-3}$ and $T_{{ref}}=10.0$, respectively.

\begin{figure}[h]
  \centering

  % --------- 第一行：rho / u ----------
  \begin{subfigure}[t]{0.43\textwidth}
    \centering
    \includegraphics[width=\linewidth]{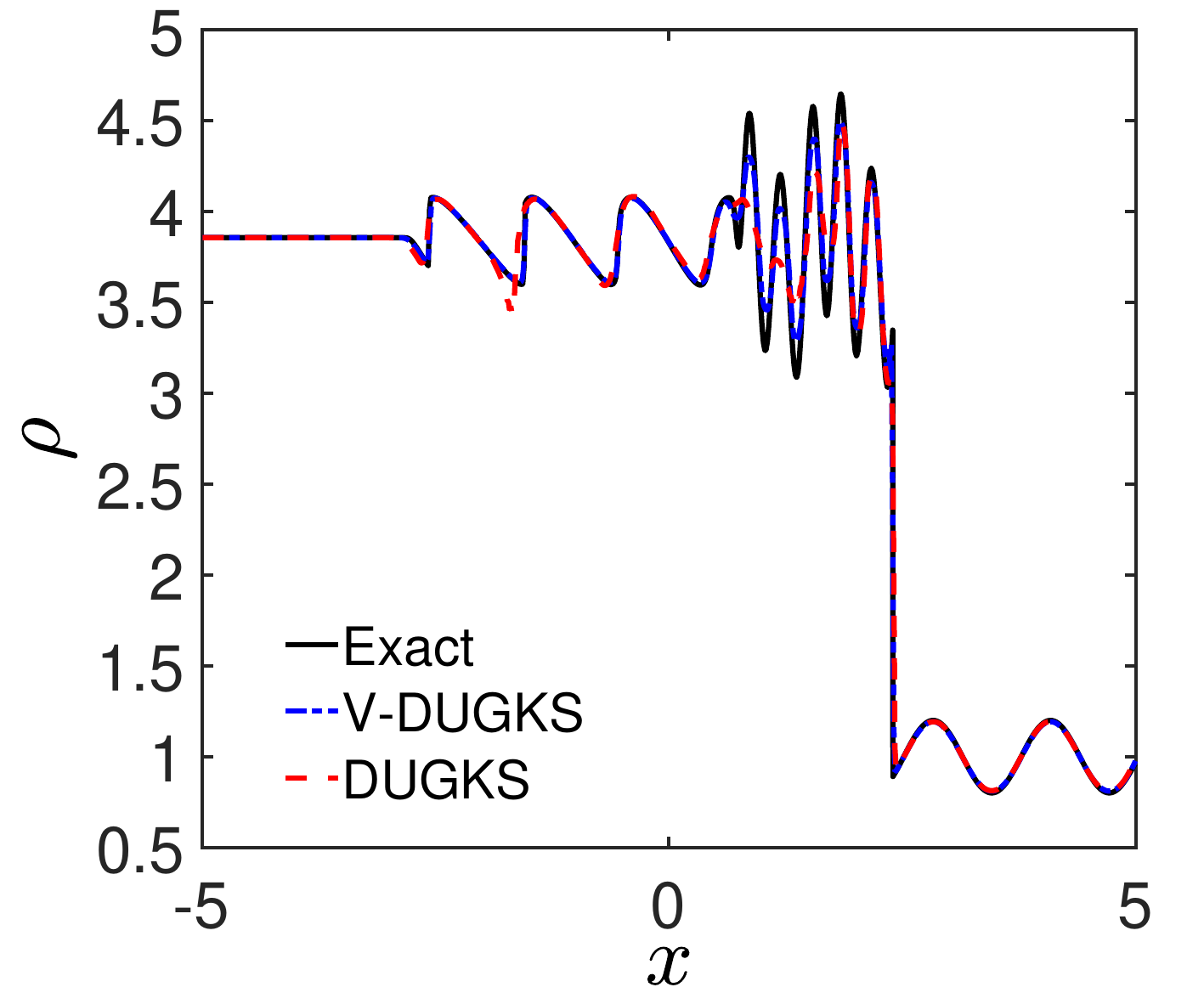}
    \caption{$\rho$}
    \label{fig:shu_osher_rho}
  \end{subfigure}\hfill
  \begin{subfigure}[t]{0.43\textwidth}
    \centering
    \includegraphics[width=\linewidth]{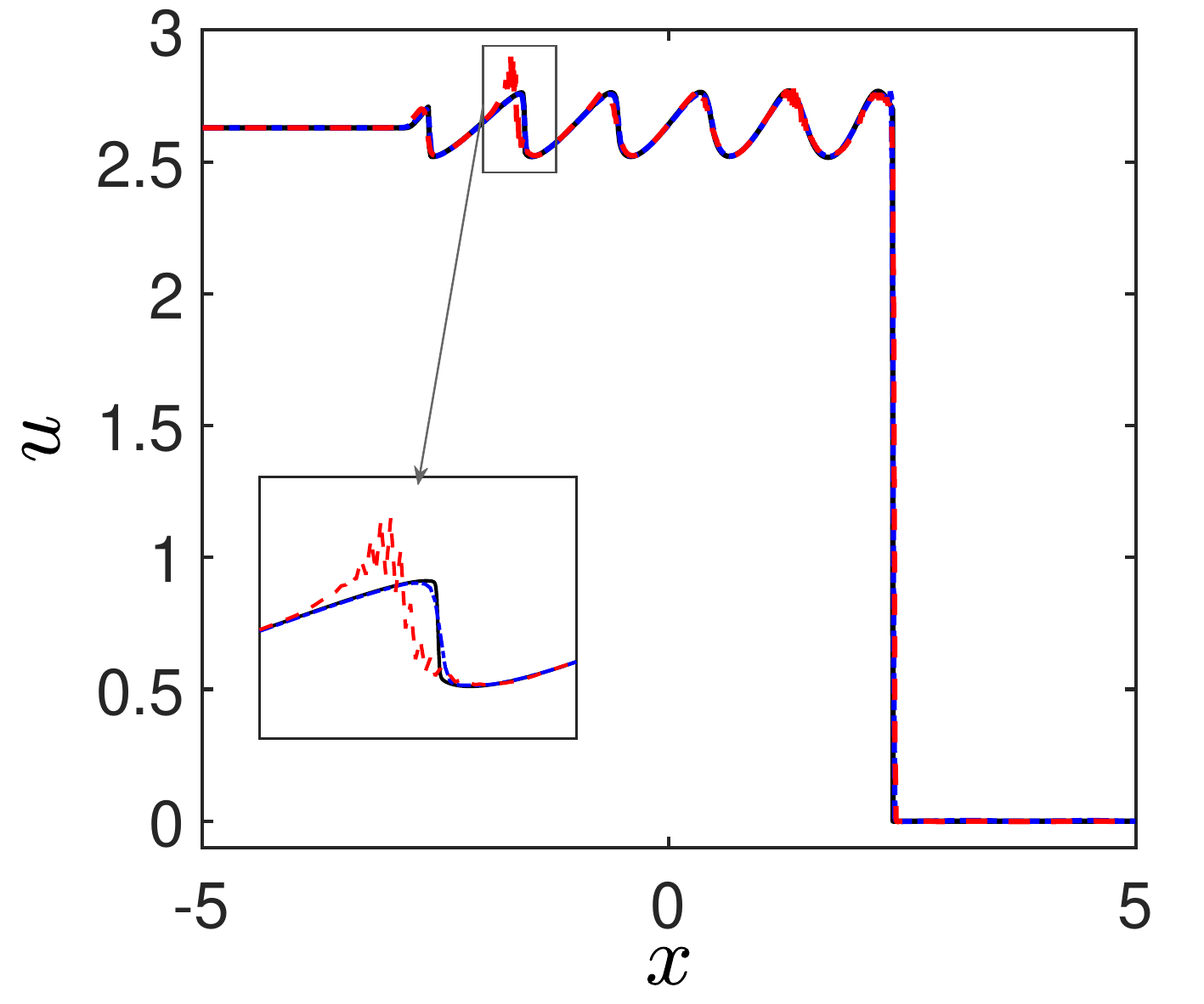}
    \caption{$u$}
    \label{fig:shu_osher_u}
  \end{subfigure}
  
  % \vspace{0.6em}

  % --------- 第二行：T / p ----------
  \begin{subfigure}[t]{0.43\textwidth}
    \centering
    \includegraphics[width=\linewidth]{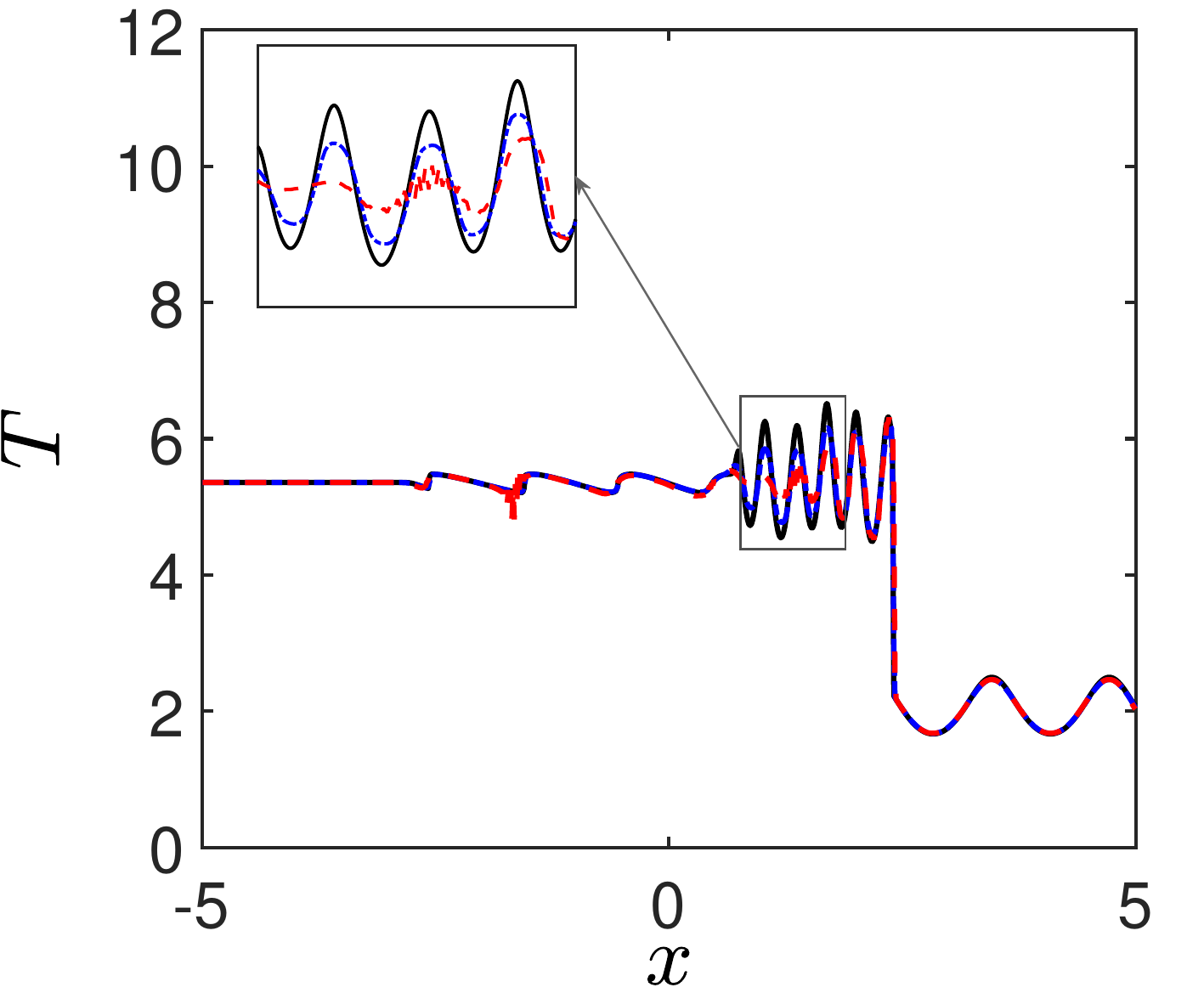}
    \caption{$T$}
    \label{fig:shu_osher_T}
  \end{subfigure}\hfill
  \begin{subfigure}[t]{0.43\textwidth}
    \centering
    \includegraphics[width=\linewidth]{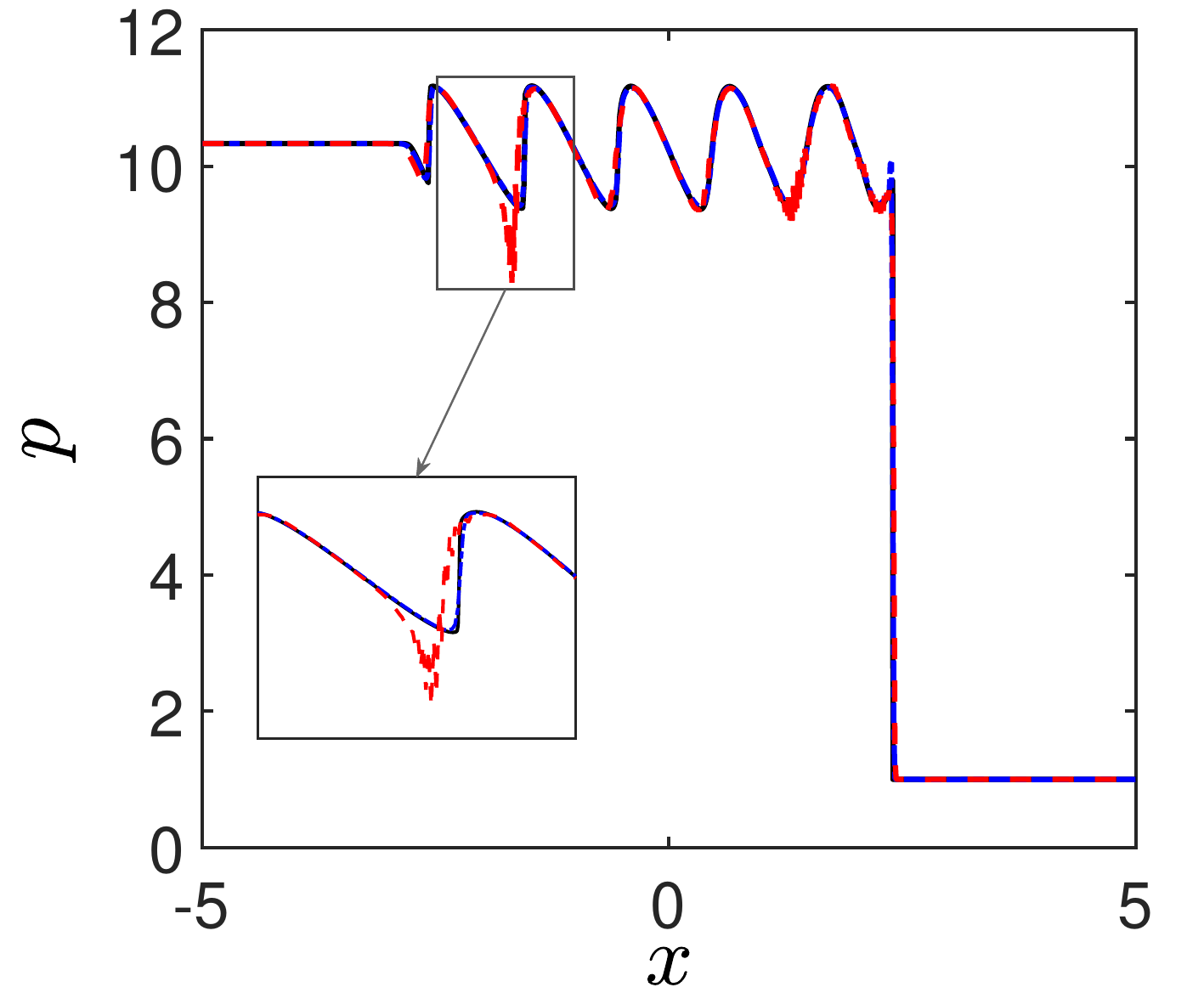}
    \caption{$p$}
    \label{fig:shu_osher_p}
  \end{subfigure}
  \caption{Comparison of density, velocity, temperature, and pressure profiles predicted by the standard DUGKS and the present V-DUGKS for the Shu--Osher problem on a uniform grid with 800 cells at $t=1.8$.}
  \label{fig:shu_osher}
\end{figure}
Figure~\ref{fig:shu_osher} displays the profiles of density, velocity, temperature, and pressure predicted by the standard DUGKS and the proposed V-DUGKS at $t=1.8$.
In addition, a reference solution obtained from a first-order scheme on a highly refined grid with $N=100000$ points is provided as reference solution.
It can be seen that both the standard DUGKS and the proposed
V-DUGKS correctly capture the main features of the Shu--Osher problem, including
the shock position and the downstream oscillatory structures.
Clear differences are observed in the post--shock region for the density and
temperature profiles.
The standard DUGKS exhibits noticeable spurious oscillations near the shock,
whereas these oscillations are significantly reduced in the V-DUGKS results,
while the physical wave patterns are preserved.
For the velocity and pressure profiles, the two schemes remain in close
agreement with the reference solution over most of the domain.
Minor discrepancies are observed in regions with strong gradients, where the
standard DUGKS shows sharper overshoots and undershoots, as highlighted in the
enlarged views.
In contrast, the V-DUGKS provides smoother transitions across the shock.
Overall, the V-DUGKS yields a more stable solution for the Shu--Osher problem,
particularly in the vicinity of the shock, without introducing excessive
numerical dissipation in the downstream oscillatory region.

\begin{figure}[htbp]
    \centering
    \begin{subfigure}[t]{0.48\textwidth}
        \centering
        \includegraphics[width=\linewidth]{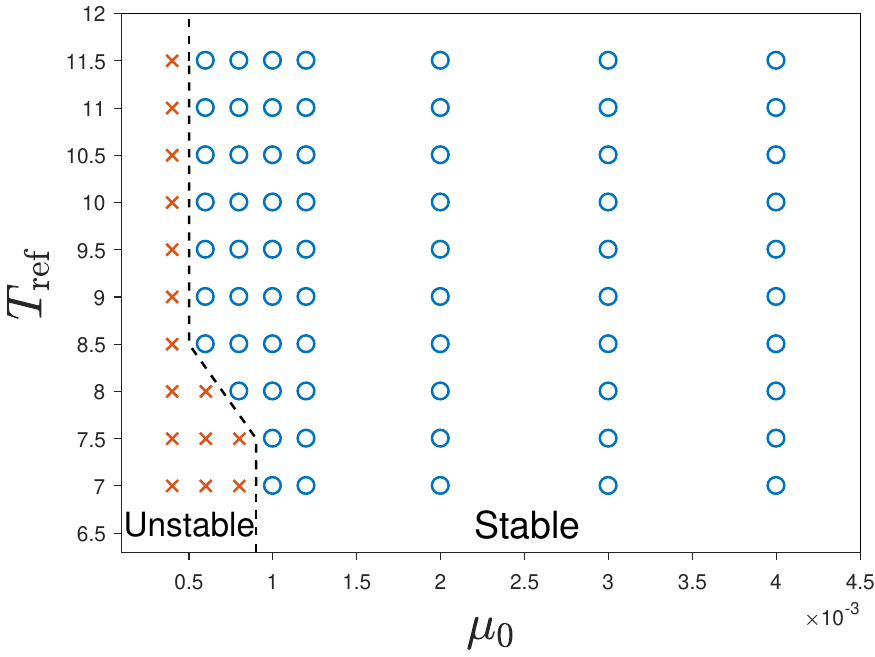}
        \caption{Stability map of the V-DUGKS.}
        \label{fig:dugks_stability}
    \end{subfigure}
    \hfill
    \begin{subfigure}[t]{0.48\textwidth}
        \centering
        \includegraphics[width=\linewidth]{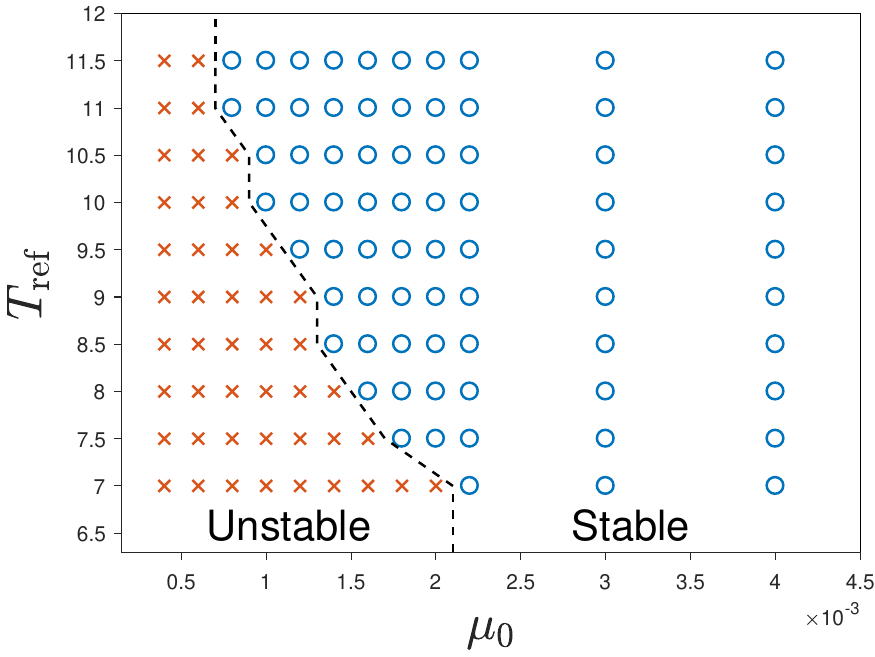}
        \caption{Stability map of the DUGKS.}
        \label{fig:V-DUGKS_stability}
    \end{subfigure}
    \caption{Stability maps of the standard DUGKS and the proposed V-DUGKS for the Shu--Osher problem over the reference viscosities $\mu_0$ and reference temperatures $T_{{ref}}$.
    ``$\circ$'' denotes stable simulations, while ``$\times$'' indicates numerical instability.}
    \label{fig:stability_comparison}
\end{figure}

Figure~\ref{fig:stability_comparison} presents the stability maps of the standard
DUGKS and the proposed V-DUGKS for the Shu--Osher problem over a range of reference
viscosities $\mu_0$ and numerical reference temperatures $T_{{ref}}$.
For each value of $\mu_0$, simulations are performed with varying
$T_{{ref}}$, and the resulting stable and unstable cases are indicated by
symbols ``$\circ$'' and ``$\times$'', respectively.
A clear difference in stability behavior can be observed between the two
schemes.
For the standard DUGKS, the stable region is confined to relatively large values
of $T_{{ref}}$, and it shrinks rapidly as the reference viscosity is
reduced.
When $\mu_0$ approaches $1.0\times10^{-3}$ and below, numerical instability
occurs over a wide range of reference temperatures, leaving only a narrow window
of stable parameter choices.
In contrast, the V-DUGKS exhibits a substantially enlarged stability region.
For moderate reference viscosities, the scheme remains stable over almost the
entire range of $T_{{ref}}$ considered.
Even at smaller values of $\mu_0$, where the standard DUGKS becomes unstable in
most cases, the V-DUGKS still preserves stability for a broad range of reference
temperatures.
These results indicate that the V-DUGKS is significantly less sensitive to the
choice of reference parameters, providing improved robustness for shock–wave
interaction problems.

\subsection{2D Riemann problem}
To quantitatively evaluate the computational efficiency of the proposed V-DUGKS, a 2D Riemann problem is considered and compared with the standard DUGKS.
The initial conditions are taken from Configuration~12 in Ref.~\cite{Riemann1998} and are given by
\begin{equation}
\label{eq:riemann2d_ic}
(\rho, u, v, p)=
\begin{cases}
(\rho_1,u_1,v_1,p_1)=(0.5313,\,0,\,0,\,0.4), & x>0.5,\; y>0.5, \\[2pt]
(\rho_2,u_2,v_2,p_2)=(1,\,0.7276,\,0,\,1),   & x\le 0.5,\; y>0.5, \\[2pt]
(\rho_3,u_3,v_3,p_3)=(0.8,\,0,\,0,\,1),     & x\le 0.5,\; y\le 0.5, \\[2pt]
(\rho_4,u_4,v_4,p_4)=(1,\,0,\,0.7276,\,1),  & x>0.5,\; y\le 0.5 .
\end{cases}
\end{equation}

The computational domain is the unit square $\Omega=[0,1]\times[0,1]$, which is uniformly discretized using a Cartesian mesh.
The time step is determined by a CFL condition~\eqref{CFL} with a fixed CFL number of $0.5$ in all simulations.
The Prandtl number and the specific heat ratio are set to $\Pr=2/3$ and $\gamma=1.4$, respectively.
A temperature--dependent viscosity model is employed, in which the dynamic viscosity is expressed as
\begin{equation}
\mu(T)=\mu_0\left(\frac{T}{T_0}\right)^{w},
\end{equation}
with $w=0.5$ and $\mu_0$ denotes the reference viscosity at the reference temperature $T_0$.
The reference density, temperature, and pressure are chosen as $\rho_0=1.0$, $T_0=1.0$, and $p_0=\rho_0 R T_0=1.0$, respectively.
Zero-gradient boundary conditions are imposed on all four boundaries, namely $\partial_{\boldsymbol{n}} f = 0$, where $\boldsymbol{n}$ denotes the outward unit normal vector.

In the present simulations, the reference viscosity is set to $\mu_0=10^{-7}$ so as to approximate the continuum regime.
With this choice, the reference mean free path $\lambda_0$ and the relaxation time $\tau_0$ are both on the order of $10^{-7}$, which are several orders of magnitude smaller than the mesh spacing and the time step dictated by the CFL condition.
Therefore, the flow under consideration lies in the continuum regime, and the comparison focuses on the computational performance rather than rarefaction effects.

For the discretization of the velocity space, the V-DUGKS and the standard DUGKS adopt different discrete velocity representations.
In the standard DUGKS, the kinetic equations are discretized using two DVSs, namely D2V12--A7 and D2V7--A5, resulting in a total of $12+7$ discrete velocities.
In contrast, the proposed V-DUGKS employs a vectorial formulation, in which four distribution components are each discretized on a D2V7--A5 set, leading to $4\times 7$ discrete velocities in total. Several DVSs satisfying the requirements are listed in Table~\ref{tab:DVS}.
As a result, the V-DUGKS employs more discrete velocity points than the standard DUGKS.
However, the difference in velocity discretization does not directly translate into a higher overall computational cost.
\begin{table}[htbp]
\centering
\setlength{\tabcolsep}{12pt} % 默认约 6pt，可按需调大
\caption{Computational time comparison between standard DUGKS and V-DUGKS for the 2D Riemann problem after 10{,}000 steps.}
\label{tab:cpu_time_10000steps}
\begin{tabular}{lcccc}
\toprule
Method & N & $400\times400$ & $800\times800$ & $1200\times1200$ \\
\midrule
\multirow{2}{*}{DUGKS}
 & Total CPU time (s)  &    1354.28     &    5491.37     &     12626.17     \\
 & Time per step (s)   &    0.135       &    0.549       &     1.263       \\
\midrule
\multirow{2}{*}{V-DUGKS}
 & Total CPU time (s)  &    1345.39     &   5594.95    &     12599.77    \\
 & Time per step (s)   &    0.135       &   0.559      &       1.260        \\
\bottomrule
\end{tabular}
\end{table}

Table~\ref{tab:cpu_time_10000steps} reports the total CPU time and the averaged computational cost per time step for both methods after 10{,}000 time steps on three different mesh resolutions.
It can be observed that the per--step computational times of the V-DUGKS and standard DUGKS are of the same order on all meshes.
In particular, although V-DUGKS employs more discrete velocities, its per--step cost remains comparable to that of the standard DUGKS, indicating that the vectorial formulation does not introduce a significant overhead in a single time step.
% \begin{table}[htbp]
% \centering
% \setlength{\tabcolsep}{12pt} % 默认约 6pt，可按需调大
% \caption{Computational time comparison between the standard DUGKS and the V-DUGKS for the 2D Riemann problem up to $t=0.25$.}
% \label{tab:cpu_time_t025}
% \begin{tabular}{lcccc}
% \toprule
% Method &  N   &  $400\times400$  & $800\times800$ & $1200\times1200$ \\
% \midrule
% DUGKS  &  Total CPU time (s)   &   149.54     &   1233.80    &   4306.58     \\
% V-DUGKS &  Total CPU time (s)   &   128.71     &   1086.92    &   3706.38      \\
% \bottomrule
% \end{tabular}
% \end{table}
\begin{table}[htbp]
\centering
\caption{Computational time comparison between the standard DUGKS and the V-DUGKS for the 2D Riemann problem up to $t=0.25$.}
\label{tab:cpu_time_t025}
\begin{tabular}{cccc}
\toprule
Mesh size & Method & Total CPU time (s) & Speed--up factor \\
\midrule
\multirow{2}{*}{$400\times400$}
 & DUGKS   & 149.54  & -- \\
 & V-DUGKS & 128.71  & 1.16 \\
\midrule
\multirow{2}{*}{$800\times800$}
 & DUGKS   & 1233.80 & -- \\
 & V-DUGKS & 1086.92 & 1.14 \\
\midrule
\multirow{2}{*}{$1200\times1200$}
 & DUGKS   & 4306.58 & -- \\
 & V-DUGKS & 3706.38 & 1.16 \\
\bottomrule
\end{tabular}
\end{table}

Since the computational efficiency of a time--marching scheme is ultimately determined by the wall--clock time required to reach a given physical time, Table~\ref{tab:cpu_time_t025} further compares the total CPU time needed by the two methods to advance the solution up to $t=0.25$.
It is observed that the V-DUGKS consistently requires less total CPU time than the standard DUGKS on all mesh resolutions considered, corresponding to a speed--up factor of approximately $1.14\text{--}1.16$.
This reduction is mainly due to the larger allowable time step in V-DUGKS under the same CFL constraint. Compared with the standard DUGKS, the compact DVS employed in the V-DUGKS has a smaller maximum discrete velocity, which reduces the number of time steps required to reach the same physical time. Meanwhile, the second--order Hermite equilibrium formulation also simplifies the evaluation of equilibrium distributions, so that the additional cost introduced by the vectorial distribution functions is largely compensated.

\begin{figure}[htbp]
    \centering
    \begin{subfigure}[b]{0.49\textwidth}
        \centering
        \includegraphics[width=\textwidth]{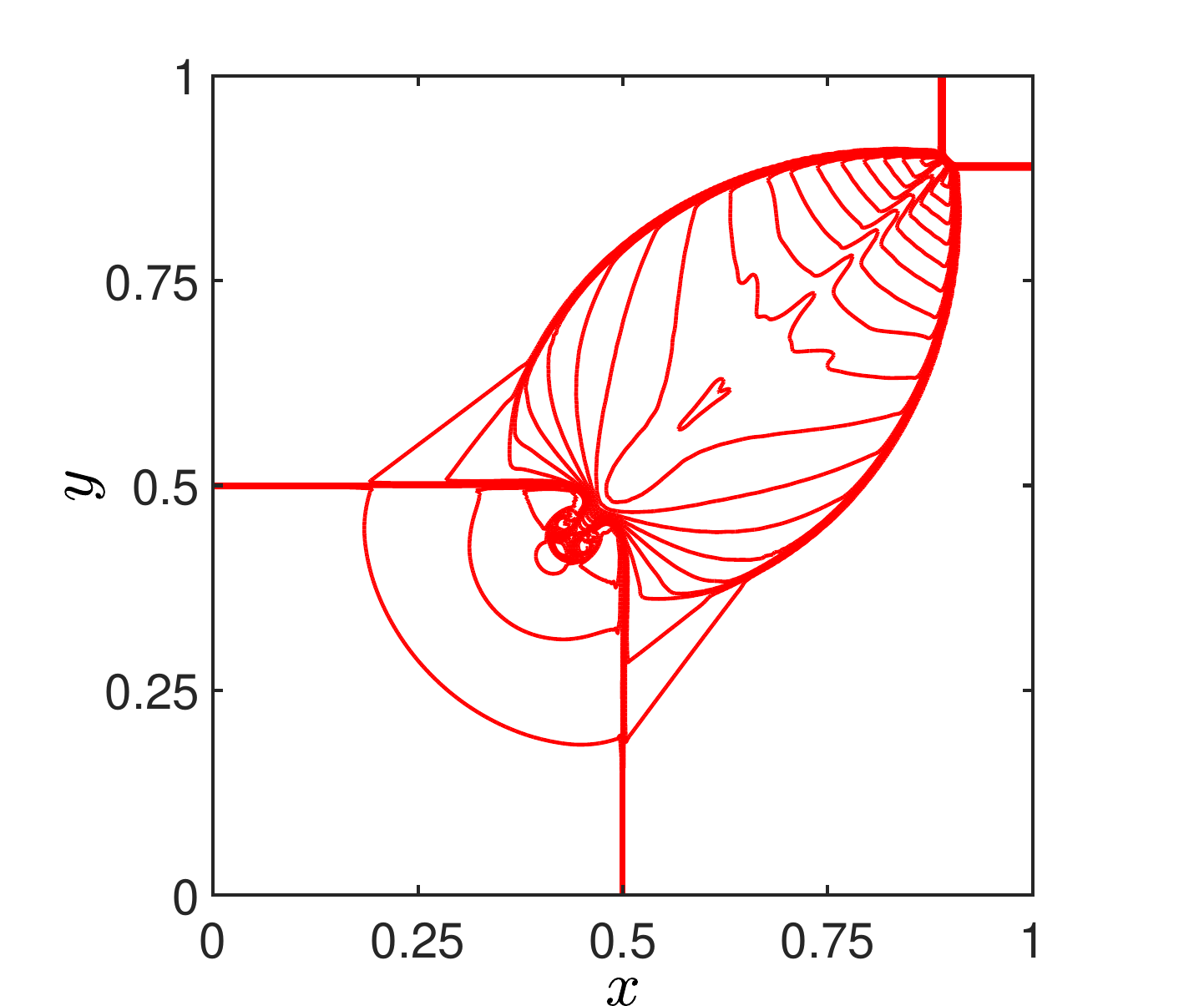}
        \caption{Density field $\rho$}
        \label{fig:riemann2d_dugks_rho_400}
    \end{subfigure}
    \hfill
    \begin{subfigure}[b]{0.49\textwidth}
        \centering
        \includegraphics[width=\textwidth]{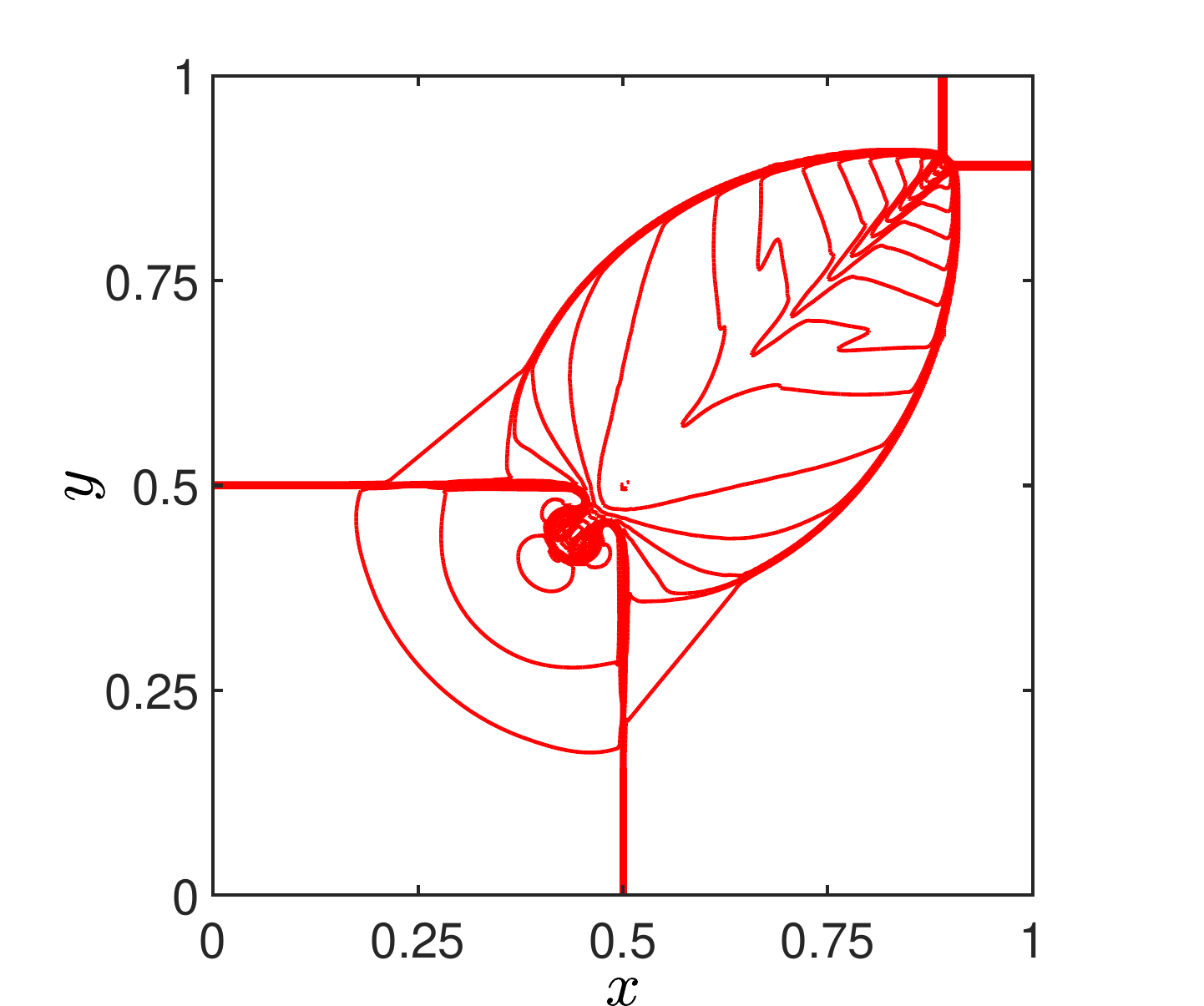}
        \caption{Temperature field $T$}
        \label{fig:riemann2d_dugks_T_400}
    \end{subfigure}

    \caption{Density and temperature fields of the 2D Riemann problem predicted by the DUGKS at $t=0.25$ on a $400\times400$ mesh.}
    \label{fig:riemann2d_dugks_400}
\end{figure}

\begin{figure}[htbp]
    \centering
    \begin{subfigure}[b]{0.49\textwidth}
        \centering
        \includegraphics[width=\textwidth]{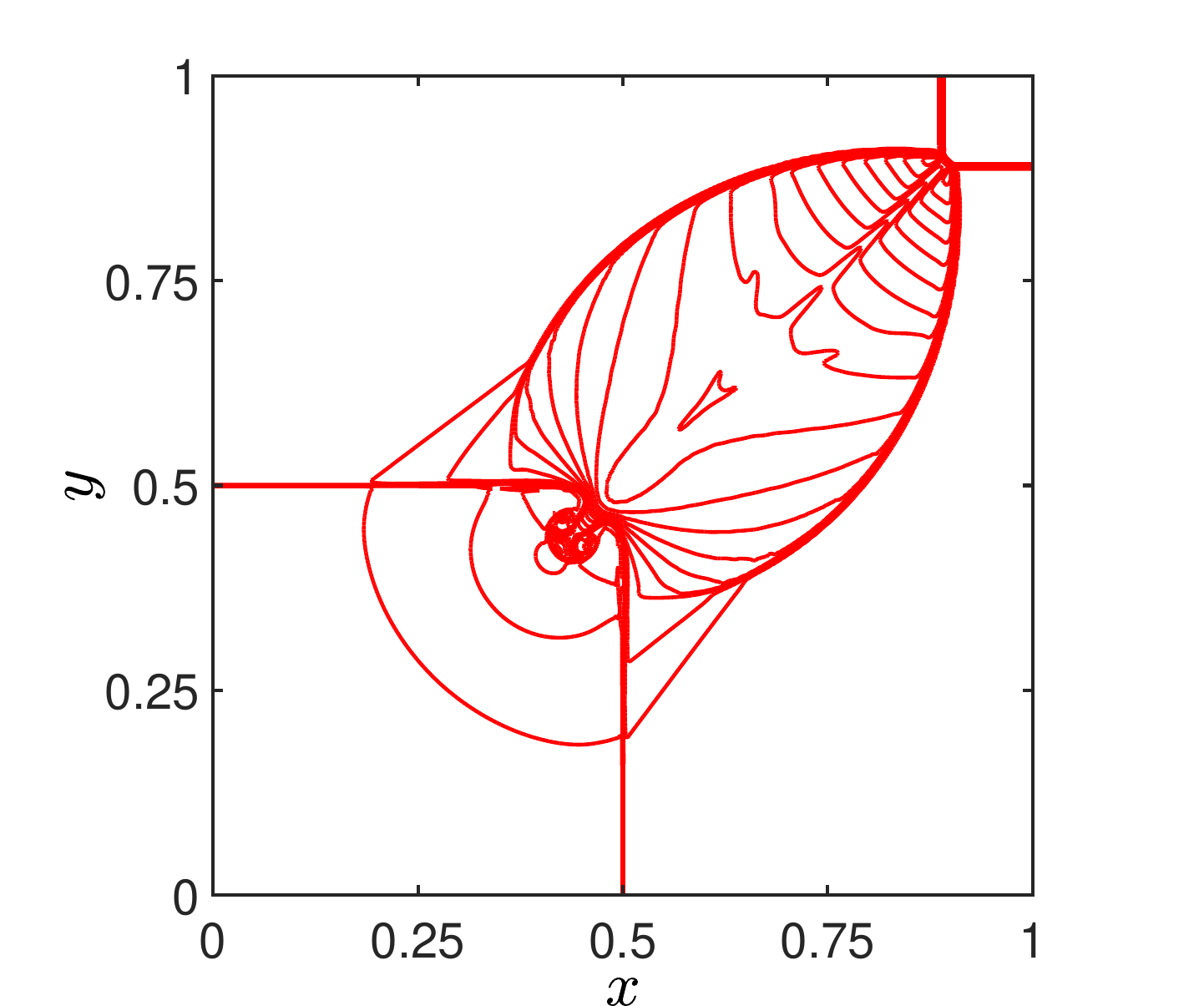}
        \caption{Density field $\rho$}
        \label{fig:riemann2d_V-DUGKS_rho_400}
    \end{subfigure}
    \hfill
    \begin{subfigure}[b]{0.49\textwidth}
        \centering
        \includegraphics[width=\textwidth]{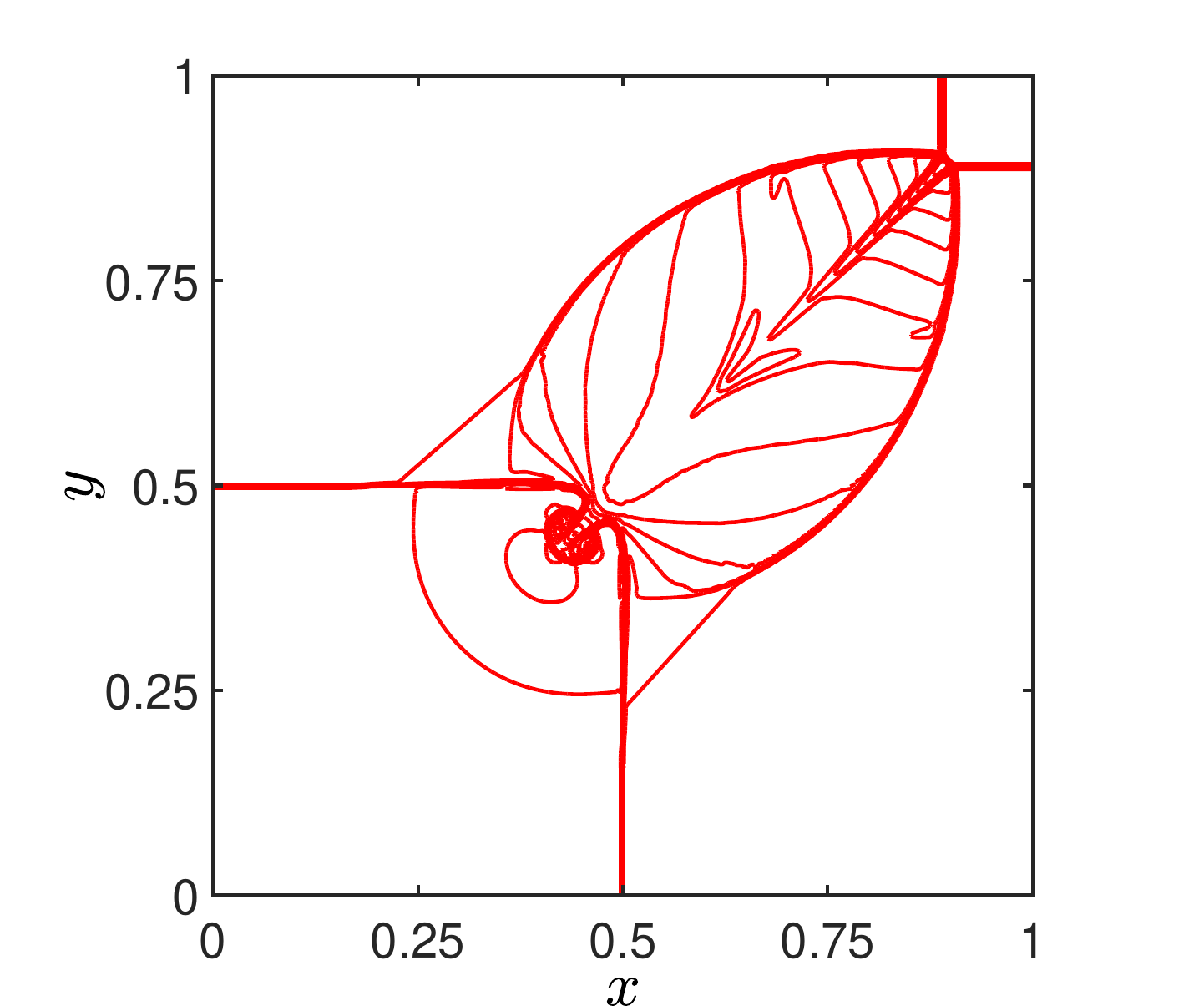}
        \caption{Temperature field $T$}
        \label{fig:riemann2d_V-DUGKS_T_400}
    \end{subfigure}

    \caption{Density and temperature fields of the 2D Riemann problem predicted by the V-DUGKS at $t=0.25$ on a $400\times400$ mesh.}
    \label{fig:riemann2d_V-DUGKS_400}
\end{figure}

For the case shown at $t=0.25$ on the $400\times400$ mesh, the density and temperature obtained by the DUGKS in Figs.~\ref{fig:riemann2d_dugks_400} vary within the ranges $\rho \in [0.5313,\,1.7189]$ and $T \in [0.7529,\,1.3834]$, respectively, while the corresponding results produced by the V-DUGKS in Figs.~\ref{fig:riemann2d_V-DUGKS_400} are $\rho \in [0.5313,\,1.7194]$ and $T \in [0.7529,\,1.3966]$.
The close agreement between the two sets of results further indicates that both methods resolve essentially the same flow structures.
Moreover, the overall wave patterns and interaction structures are consistent with those reported in previous studies~\cite{Riemann1998,Guo2013DUGKS,Guo2015DUGKS}, confirming the validity of the present simulations.

Overall, the above results demonstrate that, although the V-DUGKS employs a larger number of discrete velocity points, its computational efficiency is comparable to, and in terms of total simulation time even better than, that of the standard DUGKS.
On the one hand, the reduction in the number of required time steps to reach a given physical time effectively improves the overall efficiency.
On the other hand, the simplified equilibrium formulation in the V-DUGKS, in which third--order Hermite terms are no longer involved, reduces the per--step computational complexity.
As a result, the additional cost associated with the increased number of discrete velocity points is largely compensated.
These results confirm that the proposed V-DUGKS provides an efficient alternative to the conventional DUGKS for continuum--regime simulations, particularly for large--scale problems.

\subsection{3D Taylor--Green vortex}
As the final example, we consider the 3D Taylor--Green vortex~\cite{1937TGV} in the viscous regime to further examine the performance of the proposed V-DUGKS for the NS equations.
% The computational domain is a periodic cube $\Omega=[0,2\pi L]^3$, and all
% simulations are carried out under identical physical parameters.
% The flow is initialized using a constant--temperature initial condition (CTIC).
% The initial velocity field is prescribed as
The computational domain is a periodic box with length $L$, i.e., $\Omega=[0,2\pi L]^3$ and the initial conditions are given by
% \begin{align}
% u(x,y,z,0) &= u_0 \sin\!\left(\frac{x}{L}\right)
%              \cos\!\left(\frac{y}{L}\right)
%              \cos\!\left(\frac{z}{L}\right), \notag \\
% v(x,y,z,0) &= -u_0 \cos\!\left(\frac{x}{L}\right)
%               \sin\!\left(\frac{y}{L}\right)
%               \cos\!\left(\frac{z}{L}\right), \\
% w(x,y,z,0) &= 0, \notag
% \end{align}
% and the density field is weakly perturbed as
% \begin{equation}
% \rho(x,y,z,0)
% = \rho_0 \left\{
% 1 + \frac{C}{16}
% \left[\cos\!\left(\frac{2x}{L}\right)
%      +\cos\!\left(\frac{2y}{L}\right)\right]
% \left[\cos\!\left(\frac{2z}{L}\right)+2\right]
% \right\},
% \end{equation}
\begin{equation}
\begin{aligned}
u(x,y,z,0) &= u_0 \sin(x^*) \cos(y^*) \cos(z^*), \\
v(x,y,z,0) &= -u_0 \cos(x^*) \sin(y^*) \cos(z^*), \\
w(x,y,z,0) &= 0, \\
\rho(x,y,z,0) &= \rho_0 \left\{ 1 + \frac{C}{16} \big[\cos(2x^*) + \cos(2y^*)\big]\big[\cos(2z^*) + 2\big] \right\}, \\
T(x,y,z,0) &= T_0,\quad p(x,y,z,0) = \rho R T_0, 
\end{aligned}
\label{eq:3DTGV_IC}
\end{equation}
where $x^*=x/L$, $y^*=y/L$, $z^*=z/L$, and the variables with subscript “0” represent the constant reference values.
% the temperature is initially uniform, $T(x,y,z,0)=T_0$, and the pressure is determined from the ideal--gas equation of state, $p=\rho R T_0$.
% Here, the subscript ``$0$'' denotes reference quantities.
The coefficient $C$ is introduced to maintain consistency with the compressible formulation and depends on the Reynolds number.
Specifically, $C=1$ is adopted for $\mathrm{Re}=100$ and $\mathrm{Re}=400$, whereas for the higher--Reynolds--number case $\mathrm{Re}=1600$, $C$ is set to $\gamma \mathrm{Ma}^2$, as used in Ref.~\cite{Dominik2021semi}.
The Mach and Reynolds numbers are defined as
\begin{equation}
\mathrm{Ma}=\frac{u_0}{\sqrt{\gamma R T_0}},
\qquad
\mathrm{Re}=\frac{\rho_0 L u_0}{\mu_0},
\end{equation}
with $\rho_0$ and $\mu_0$ being the reference density and dynamic viscosity, respectively.

In all simulations, the parameters are fixed as
$L=1.0$, $T_0=1.0$, $u_0=1.0$, and $\rho_0=1.0$.
The computational domain is discretized using a uniform mesh of $256^3$ control volumes unless otherwise stated.
Accordingly, the gas constant $R$ and the reference viscosity $\mu_0$ are determined by the prescribed Mach and Reynolds numbers.
The ratio of specific heats is taken as $\gamma=1.4$, and the Prandtl number is set to $\mathrm{Pr}=0.71$.
The local viscosity follows the Sutherland's law~\cite{Sutherland2018,Sutherland2021},
\begin{equation}
\mu
= \mu_0
\frac{1.4042\,(T/T_0)^{1.5}}{T/T_0+0.40417}.
\end{equation}

To discretize the velocity space, the standard DUGKS~\cite{GUO2023Compressible} adopts two different DVSs for the two distribution functions: the D3V27--A7 and the D3V13--A5 in table~\ref{tab:DVS}, resulting in $27+13$ discrete velocities in total.
In contrast, the proposed V-DUGKS employs the D3V13--A5 for each of the five distribution functions, leading to a total $5\times 13$ discrete velocity points.
The discrete distributions are initialized by adding first--order Hermite polynomial expansion,
\begin{equation}
f_{\alpha,i} = f_{\alpha,i}^{(eq)} + f_{\alpha,i}^{(neq)}, \qquad
\alpha=0,1,...,4,
\end{equation}
where the nonequilibrium parts are expressed in terms of the viscous stress tensor and the heat flux as
\begin{equation}
\label{eq:ce_init_neq}
f_{\alpha,i}^{(neq)}=
\begin{cases}
\displaystyle
0, \qquad \alpha = 0, \\[0.2ex]
\displaystyle
 -\,w_i\,
\frac{\boldsymbol{\sigma} \cdot \! \boldsymbol{\xi}_i}
{R T_0}, \qquad \alpha = 1,2,3,  \\[1.2ex]
\displaystyle
 -\,w_i\,
\frac{\left(\kappa \nabla T + \boldsymbol{\sigma}\cdot\boldsymbol{u}\right)\cdot \boldsymbol{\xi}_i}
{R T_0}, \qquad \alpha = 4,
\end{cases}
\end{equation}
with $w_i$ the quadrature weights associated with the selected DVS,
the viscous stress tensor $\boldsymbol{\sigma}$ and the temperature gradient
$\nabla T$ evaluated from the prescribed initial conditions.

The evolution of the flow is characterized by several integral quantities, including the total kinetic energy $E$, the solenoidal dissipation rate $\varepsilon_s$, the compressive dissipation rate $\varepsilon_c$, and the total dissipation $\varepsilon$, which are defined as
\begin{equation}
\begin{aligned}
E &= \frac{1}{\rho_0 u_0^2 V}
     \int_{\Omega} \frac{1}{2}\rho\,|\boldsymbol{u}|^2\,{d}\boldsymbol{x}, \qquad
\varepsilon_s =
\frac{1}{\mathrm{Re}\,\rho_0 u_0^2 \mu_0 V}
\int_{\Omega} \mu\,|\nabla\times\boldsymbol{u}|^2\,{d}\boldsymbol{x}, \\
\varepsilon_c &=
\frac{1}{\mathrm{Re}\,\rho_0 u_0^2 \mu_0 V}
\int_{\Omega}
\left(\mu_b+\frac{4}{3}\mu\right)
(\nabla\cdot\boldsymbol{u})^2
\,{d}\boldsymbol{x}, \qquad
\varepsilon = \varepsilon_s + \varepsilon_c ,
\end{aligned}
\end{equation}
where $V$ denotes the volume of the computational domain.

\begin{figure}[htbp]
    \centering
    \begin{subfigure}[b]{0.48\textwidth}
        \centering
        \includegraphics[width=\textwidth]{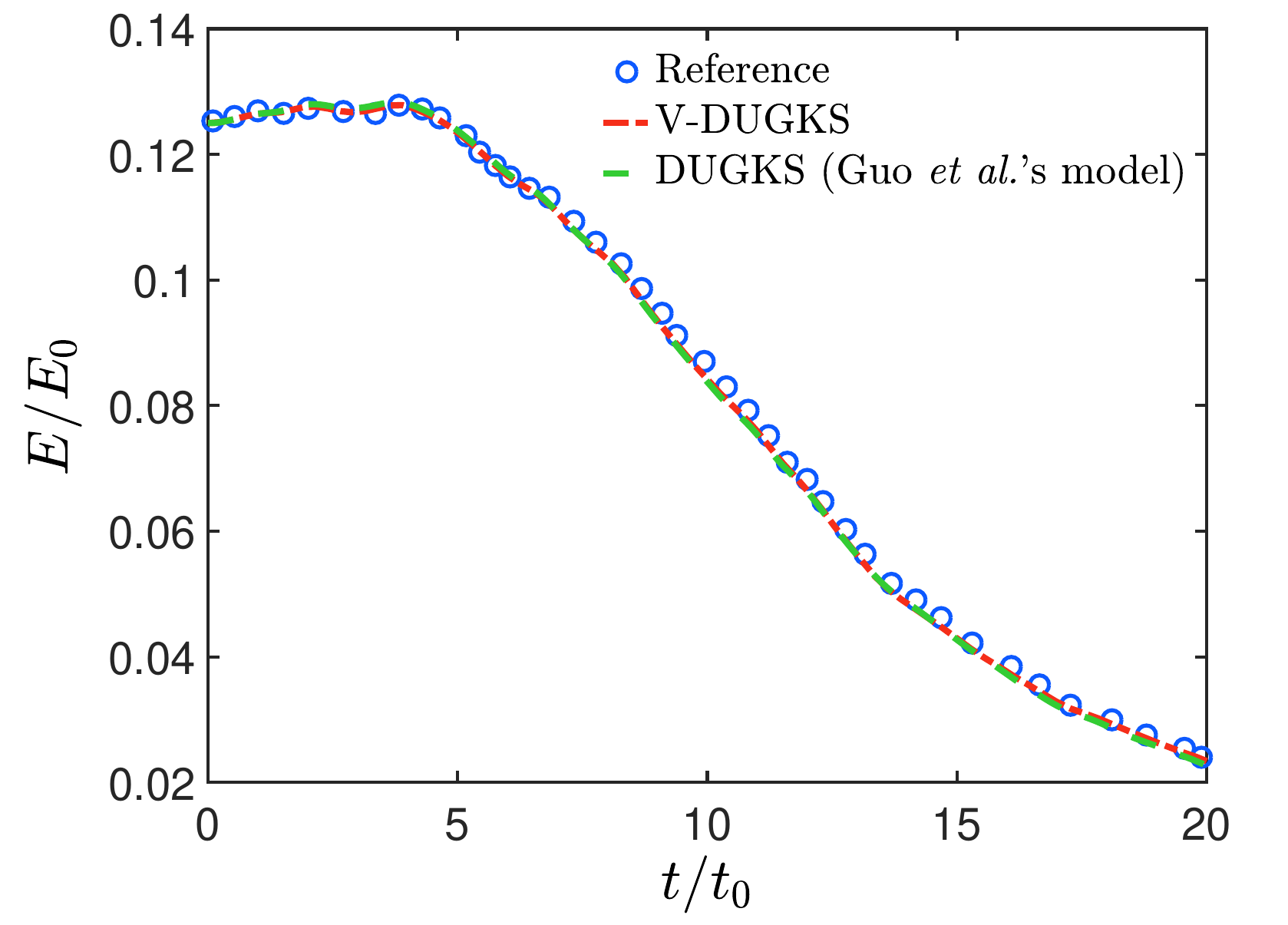}
        \caption{Total kinetic energy}
        \label{fig:tgv_energy_1_25}
    \end{subfigure}
    \hfill
    \begin{subfigure}[b]{0.48\textwidth}
        \centering
        \includegraphics[width=\textwidth]{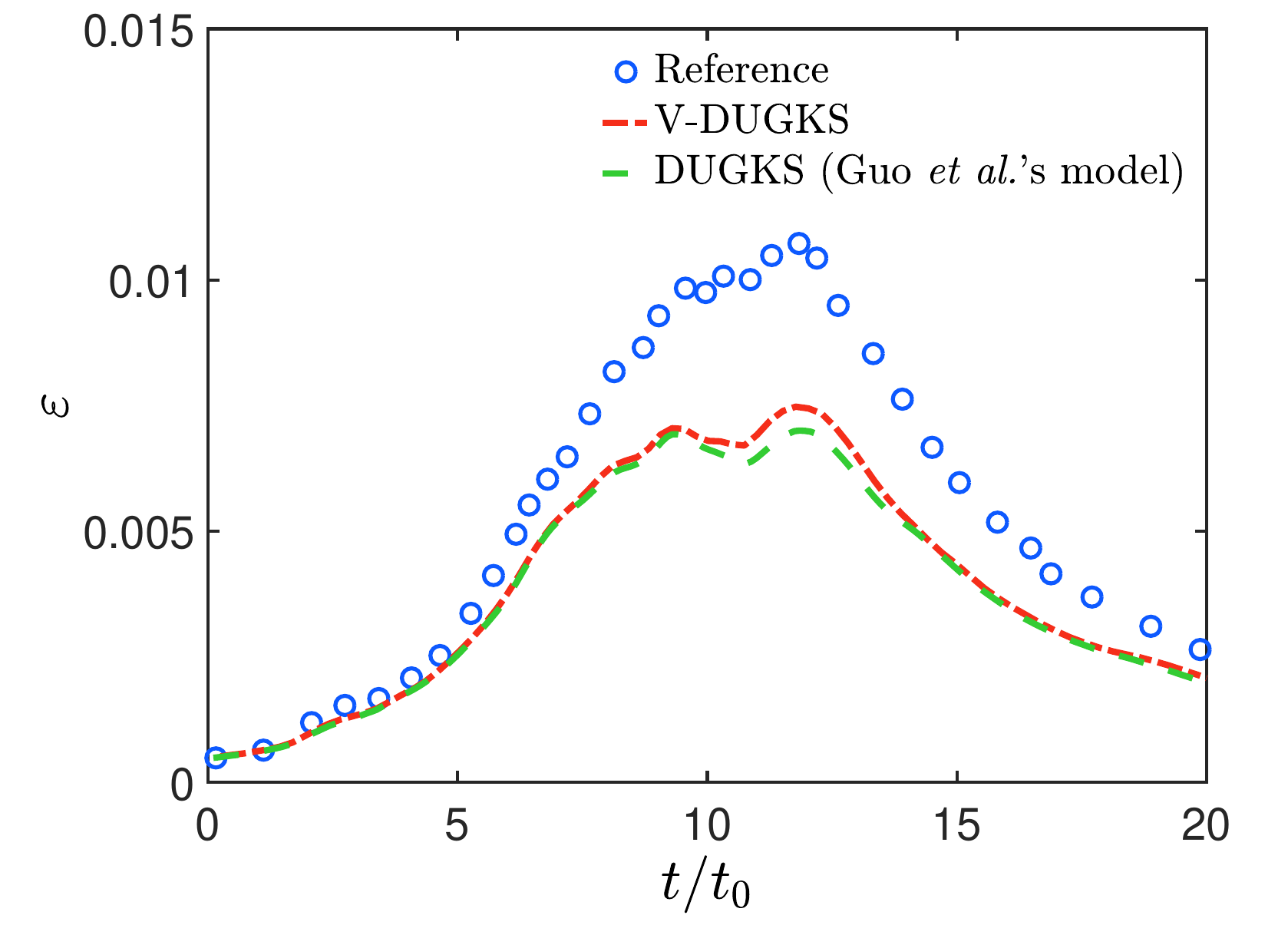}
        \caption{Total dissipation rate}
        \label{fig:tgv_dissipation_1_25}
    \end{subfigure}
    \caption{Temporal evolution of the total kinetic energy (a) and total dissipation rate (b) for the 3D Taylor--Green vortex at $\mathrm{Ma}=1.25$ on a $256^3$ mesh.
    Results obtained with DUGKS~\cite{GUO2023Compressible} and the proposed V-DUGKS are compared against a reference solution taken from~\cite{Sutherland2021}.
    The reference data correspond to a sixth--order Targeted Essentially Non–Oscillatory (TENO) finite-difference method
    on a $768^3$ mesh.
    }
    \label{fig:tgv_1_25}
\end{figure}
Figure~\ref{fig:tgv_1_25} presents the time evolution of the total kinetic
energy and the total dissipation rate for the 3D Taylor--Green vortex at $\mathrm{Ma}=1.25$.
It can be observed that both standard DUGKS and V-DUGKS are able to capture the overall decay of kinetic energy and the growth and subsequent attenuation of the
dissipation rate.
% However, the V-DUGKS exhibits improved robustness and better agreement with the
% reference solution, especially around the peak dissipation stage.
Nevertheless, noticeable differences appear in the dissipation history,
especially near the peak dissipation stage, where strong nonlinear interactions
and intensive small--scale structures place higher demands on numerical
stability.
In this regime, the V-DUGKS exhibits a smoother evolution and closer agreement
with the reference solution, while the standard DUGKS shows relatively larger
deviations.
% The improved performance of the V-DUGKS is mainly attributed to its enhanced
% robustness in strongly compressible flows, leading to a more stable prediction
% of the dissipation dynamics in the supersonic 3D TGV.
It is worth noting that, as reported in Ref.~\cite{GUO2023Compressible}, the D3V27--A7 set may encounter stability issues for the 3D
Taylor--Green vortex at $\mathrm{Ma}=1.25$.
In the present study, such difficulties are effectively avoided by adopting the
D3V13--A5 set uniformly in the V-DUGKS formulation.
This choice leads to a more stable evolution of both kinetic energy and
dissipation rate, demonstrating the enhanced numerical stability of the proposed
scheme in the supersonic regime.

\begin{figure}[htbp]
    \centering
    \begin{subfigure}[b]{0.48\textwidth}
        \centering
        \includegraphics[width=\textwidth]{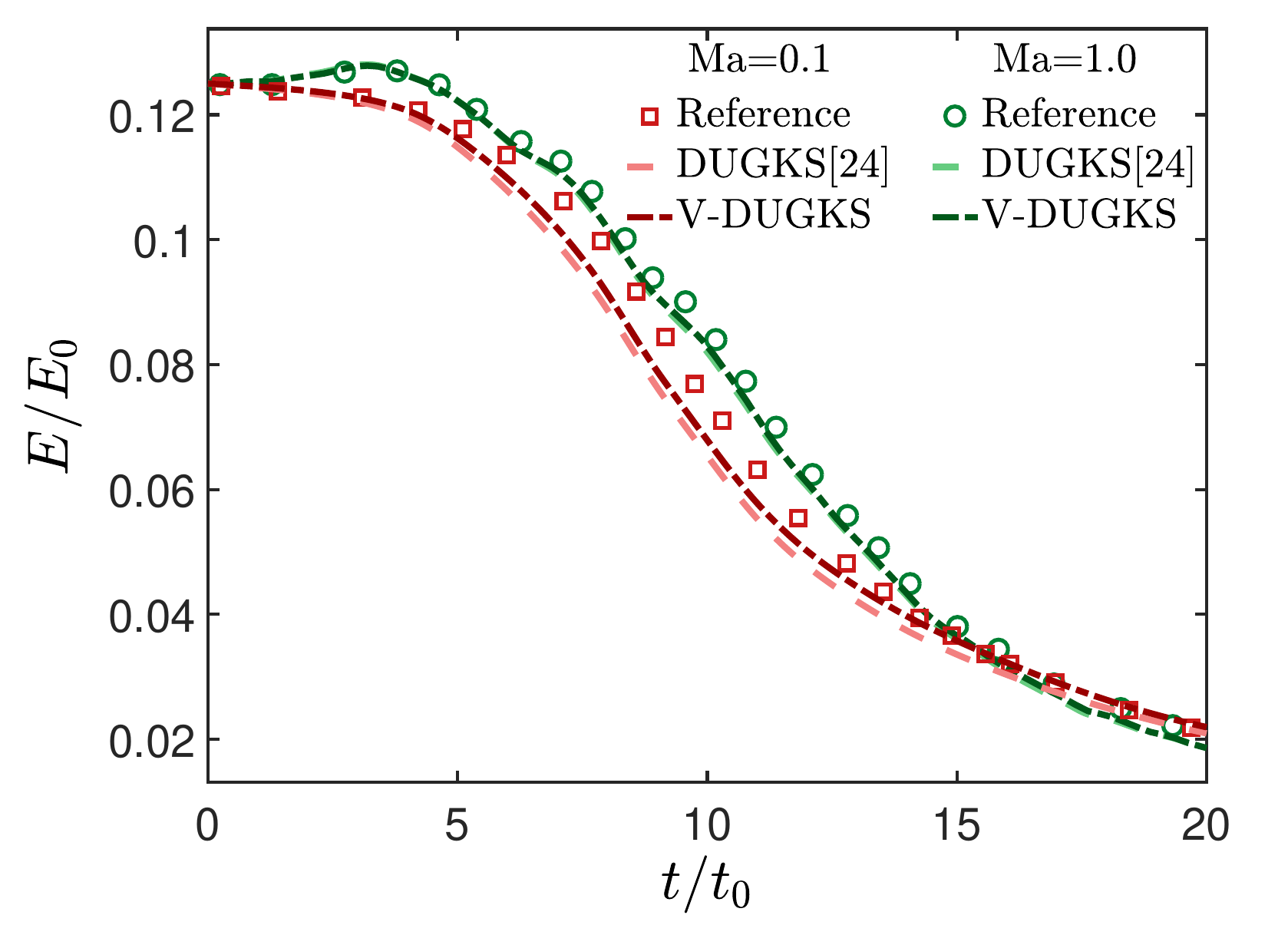}
        \caption{Total kinetic energy}
        \label{fig:tgv_energy_1_0}
    \end{subfigure}
    \hfill
    \begin{subfigure}[b]{0.48\textwidth}
        \centering
        \includegraphics[width=\textwidth]{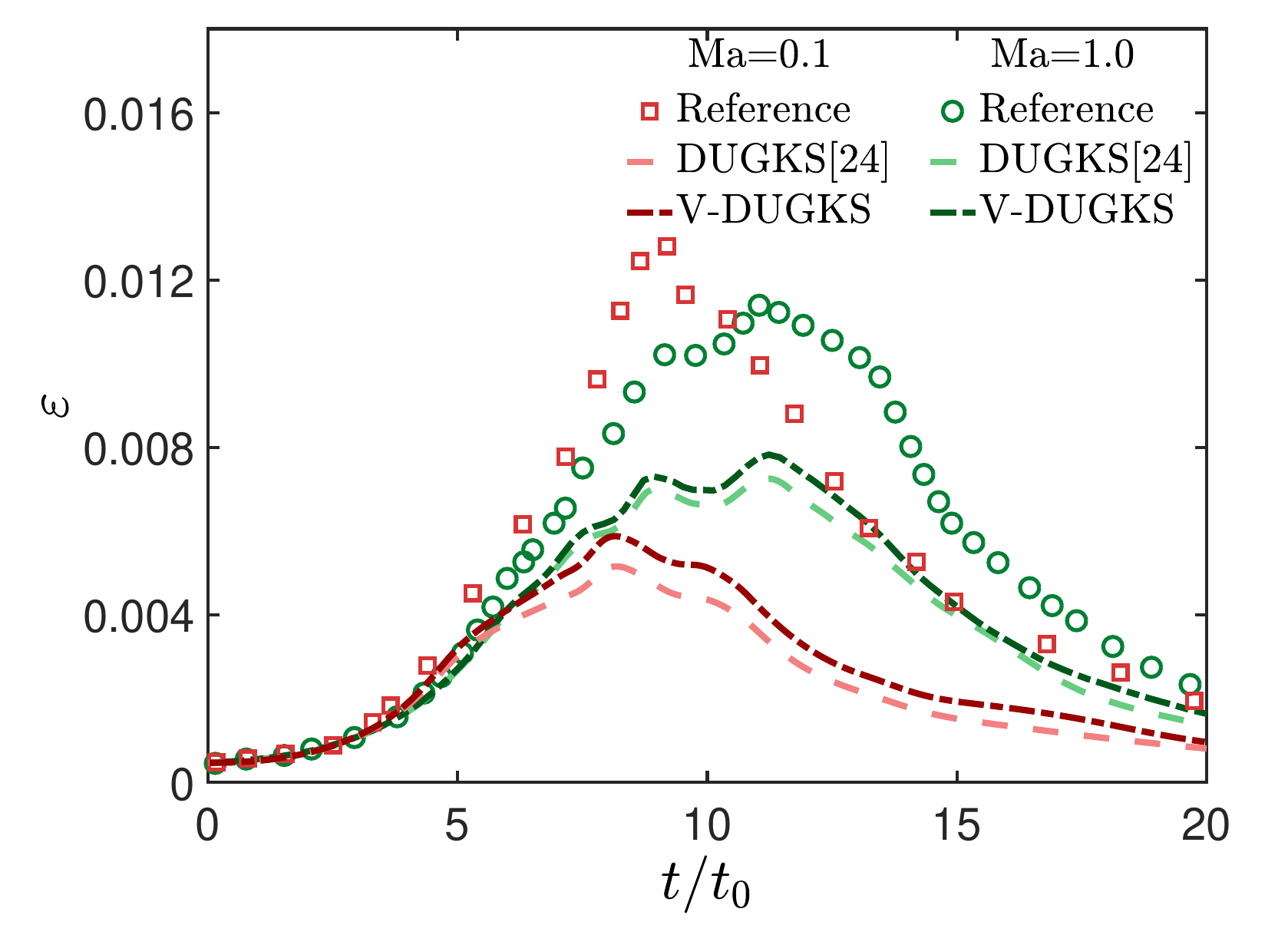}
        \caption{Total dissipation rate}
        \label{fig:tgv_dissipation_1_0}
    \end{subfigure}
    \caption{Time histories of the total kinetic energy (a) and total dissipation
    rate (b) for the 3D Taylor--Green vortex under subsonic
    ($\mathrm{Ma}=0.1$) and transonic ($\mathrm{Ma}=1.0$) conditions.
    Results obtained with DUGKS~\cite{GUO2023Compressible} and V-DUGKS are
    compared with the corresponding reference solutions~\cite{Sutherland2021,Wang2013}.
    The reference data for $\mathrm{Ma}=0.1$ are taken from a spectral solution
    on a $512^3$ mesh, while those for $\mathrm{Ma}=1.0$ are obtained using a
    sixth--order TENO scheme on a $1024^3$ mesh.}
    \label{fig:tgv_1_0}
\end{figure}

\begin{table}[htbp]
\centering
\caption{The total CPU time running to $t=20$ at different Mach numbers on a uniform mesh of $256^3$ cells.}
\label{tab:tgv3d_cpu_time}
\begin{tabular}{cccc}
\toprule
Mach number & Method & Total CPU time (h) & Speed--up factor \\
\midrule
\multirow{2}{*}{$\mathrm{Ma}=0.1$}
 & DUGKS   & 275.68 & -- \\
 & V-DUGKS & 96.45  & $2.86$ \\
\midrule
\multirow{2}{*}{$\mathrm{Ma}=1.0$}
 & DUGKS   & 31.78 & -- \\
 & V-DUGKS & 13.46 & $2.36$ \\
\midrule
\multirow{2}{*}{$\mathrm{Ma}=1.25$}
 & DUGKS   & 26.54 & -- \\
 & V-DUGKS & 11.83 & $2.24$ \\
\bottomrule
\end{tabular}
\end{table}

To further assess the performance of the proposed scheme in different flow
regimes, Fig.~\ref{fig:tgv_1_0} compares the time histories of the total kinetic
energy and total dissipation rate at $\mathrm{Ma}=0.1$ and $\mathrm{Ma}=1.0$.
For both cases, the overall decay of kinetic energy is captured by the two
schemes and remains close to the reference solutions.
A closer inspection shows that the V-DUGKS follows the reference energy curve
slightly better over the main decay interval, while the standard DUGKS tends to
exhibit a larger deviation.
More noticeable differences are observed in the dissipation histories.
Around the peak dissipation stage, the V-DUGKS provides a closer prediction to
the reference data, whereas the standard DUGKS underestimates the peak and shows
a larger discrepancy during the subsequent decay.
These results confirm that the proposed V-DUGKS preserves good accuracy
in subsonic and transonic regimes, with a modest but consistent improvement in
the prediction of key integral quantities.

In addition to the improved robustness, the computational efficiency of the
V-DUGKS is further examined. As shown in Table~\ref{tab:tgv3d_cpu_time},
the total CPU time required to advance the solution to $t=20$ on a uniform
$256^3$ grid is compared for different Mach numbers.
It is observed that the V-DUGKS consistently requires less computational time
than the standard DUGKS for all cases considered, corresponding to a speed--up
factor of approximately $2.2\text{--}2.9$ across the entire range of Mach numbers.
The efficiency gain is more pronounced in the low--Mach--number regime, while a
substantial speed--up is still maintained for transonic and supersonic flows.
This improvement can be mainly attributed to the larger allowable time step of
V-DUGKS under the same CFL constraint and the reduced complexity of the
second--order Hermite equilibrium formulation.
These results demonstrate that the proposed V-DUGKS achieves improved numerical
robustness together with a significant enhancement in computational efficiency for three--dimensional viscous simulations.

\section{CONCLUSIONS} \label{sec:5}
In this work, an efficient V-DUGKS has been constructed and validated for the numerical simulation of continuum compressible flows. By adopting a vectorial kinetic formulation, the transport processes of mass, momentum, and total energy are described in a unified manner, providing a stable and consistent kinetic framework for the compressible NS equations.

Numerical results demonstrate that, compared with the standard DUGKS, the V-DUGKS exhibits enhanced robustness under strongly compressible flow conditions. In problems involving shock waves, strong nonlinear interactions, and complex flow structures, the proposed scheme effectively suppresses nonphysical oscillations while maintaining the overall consistency of the solution. The improved stability is primarily reflected in the increased robustness with respect to numerical parameter choices, without relying on additional numerical dissipation or empirical corrections.

From the perspective of computational efficiency, although the vectorial formulation introduces an increased number of discrete velocity degrees of freedom, numerical experiments indicate that the computational cost per time step remains comparable to that of the standard DUGKS. This behavior can be attributed to the simplified equilibrium construction and reduced complexity of velocity-moment evaluations, which prevent a noticeable increase in per-step computational overhead. Moreover, under the same CFL constraint, the V-DUGKS permits larger stable time steps, thereby reducing the number of time-marching steps required to reach a given physical time. As a result, the overall wall-clock time is significantly reduced in multi-dimensional simulations.

In summary, the proposed V-DUGKS achieves a favorable balance between numerical stability and computational efficiency. The vectorial kinetic model offers a competitive and robust alternative for continuum-regime compressible flow simulations, particularly for this problem in which both stability and computational cost are critical considerations.

% \begin{nomenclature}
%     $A$      & = & Amplitude \\
%     $C_d$    & = & drag coefficient \\
%     $f_e$    & = & linearization coefficient \\
%     $K_i$    & = & modification factor \\
% \end{nomenclature}

\newpage
\appendix

\section{Chapman--Enskog analysis}
First, we introduce the following Chapman-Enskog (CE) expansions~\cite{chapman1970}:
\begin{subequations}
\begin{align}
\partial_t &= \sum_{n=1}^{\infty} \varepsilon^{n} \partial_{t_n}, \qquad
\nabla = \varepsilon \nabla_1, \label{A4b} \\[6pt]
R&=\varepsilon R^{(1)}, \qquad
f_\alpha = \sum_{n=0}^{\infty} \varepsilon^{n} f_\alpha^{(n)}, \qquad \alpha=0,1,...,D+1
\end{align}
\end{subequations}
where $\varepsilon$ is a small expansion parameter.
By inserting these expansions into the kinetic equations~\eqref{vec-BGK}
and equating terms of the same orders of $\varepsilon$,
a hierarchy of equations at $\varepsilon^0$, $\varepsilon^1$, and $\varepsilon^2$
is obtained for the distribution functions,
\begin{subequations}\label{f0}
\begin{align}
\varepsilon^{0}: \quad  f_0^{(0)} &= f_0^{(eq)}, \label{f0_a} \\[6pt]
\varepsilon^{1}: \quad  D_{1} f_0^{(0)} 
    &= - \frac{f_0^{(1)}}{\tau_f},  \label{f0_b}\\[6pt]
\varepsilon^{2}: \quad  \partial_{t_2} f_0^{(0)} + D_{1} f_0^{(1)}
    &= - \frac{f_0^{(2)}}{\tau_f}, \label{f0_c}
\end{align}
\end{subequations}
\vspace{-24pt}
\begin{subequations}\label{f_alpha}
\begin{align}
\varepsilon^{0}: \quad  \boldsymbol{f}^{(0)} &= \boldsymbol{f}^{(eq)},  \label{f_alpha_a}\\[6pt]
\varepsilon^{1}: \quad  D_{1} \boldsymbol{f}^{(0)} 
    &= - \frac{\boldsymbol{f}^{(1)}}{\tau_f},  \label{f_alpha_b}  \\[6pt]
\varepsilon^{2}: \quad  \partial_{t_2} \boldsymbol{f}^{(0)} + D_{1} \boldsymbol{f}^{(1)}
    &= - \frac{\boldsymbol{f}^{(2)}}{\tau_f}, \label{f_alpha_c}
\end{align}
\end{subequations}
\vspace{-24pt}
\begin{subequations}\label{f_D+1}
\begin{align}
\varepsilon^{0}: \quad  f_{D+1}^{(0)} &= f_{D+1}^{(eq)}, \label{f_D+1_a} \\[6pt]
\varepsilon^{1}: \quad  D_{1} f_{D+1}^{(0)} 
    &= - \frac{f_{D+1}^{(1)}}{\tau_e}
    + \frac{R^{(1)}}{\tau_{ef}},  \label{f_D+1_b} \\[6pt]
\varepsilon^{2}: \quad  \partial_{t_2} f_{D+1}^{(0)} + D_{1} f_{D+1}^{(1)}
    &= - \frac{f_{D+1}^{(2)}}{\tau_e}, \label{f_D+1_c}
\end{align}
\end{subequations}
where $ D_{1} = \partial_{t_1} + \boldsymbol{\xi} \cdot \nabla_1 $. Equations \eqref{f0_a}, \eqref{f_alpha_a} and \eqref{f_D+1_a}, combined with Eq.~\eqref{conservedeq} indicate that
\begin{subequations}
\begin{align}
\int f_\alpha^{(n)}\, d\boldsymbol{\xi} = 0, \quad \alpha = 0,1,...,D+1.
\end{align}
\end{subequations}

At the $\varepsilon$ order, taking the zeroth-order velocity moments of
Eqs.~\eqref{f0_b}, \eqref{f_alpha_b}, and \eqref{f_D+1_b}
yields
\begin{subequations}\label{varepsilon1}
\begin{align}
\partial_{t_1} \rho + \nabla_1 \cdot (\rho \boldsymbol{u}) = 0, \\[6pt]
\partial_{t_1} (\rho \boldsymbol{u}) + \nabla_1 \cdot (\rho \boldsymbol{u} \boldsymbol{u} + p \boldsymbol{I}) = 0, \\[6pt]
\partial_{t_1} (\rho E) + \nabla_1 \cdot [(p+ \rho E) \boldsymbol{u}] = 0,
\end{align}
\end{subequations}
where $p = \rho R T$ represents the pressure.
Following the same procedure, the zeroth-order velocity moments of
Eqs.~\eqref{f0_c}, \eqref{f_alpha_c}, and \eqref{f_D+1_c}
at the $\varepsilon^2$ order lead to
\begin{subequations}
\begin{align}
\partial_{t_2} \rho = 0,
\\[8pt]
\partial_{t_2} (\rho \boldsymbol{u})
+ \nabla_1 \cdot \boldsymbol{\Pi}^{(1)} = 0,
\\[8pt]
\partial_{t_2} (\rho E)
+ \nabla_1 \cdot \boldsymbol{Q}^{(1)} = 0.
\end{align}
\end{subequations}
where
\(
\boldsymbol{\Pi}^{(1)} = \int \boldsymbol{\xi} \boldsymbol{f}^{(1)}\, d\boldsymbol{\xi}
\)
and
\(
\boldsymbol{Q}^{(1)} = \int \boldsymbol{\xi} f_{D+1}^{(1)}\, d\xi.
\)

To recover the equations at the $\varepsilon^2$ order,
we employ the following relations obtained by transforming
Eq.~\eqref{varepsilon1}:
\begin{subequations} \label{aids}
\begin{align}
\partial_{t_1} p
+ \nabla_1 \cdot (p \boldsymbol{u})
&= -\frac{2}{K+3}\, p\, \nabla_1 \cdot \boldsymbol{u},
\label{aids_a} \\[6pt]
\partial_{t_1} (\rho\, \boldsymbol{u} \boldsymbol{u})
+ \nabla_1 \cdot (\rho\, \boldsymbol{u} \boldsymbol{u} \boldsymbol{u})
&= -[\boldsymbol{u}\, \nabla_1 p]_{s},
\label{aids_b} \\[6pt]
\partial_{t_1} (p\, \boldsymbol{u})
+ \nabla_1 \cdot (p\, \boldsymbol{u}\boldsymbol{u})
&= - RT\, \nabla_1 p
   - \frac{2}{K+3}(\nabla_1 \cdot \boldsymbol{u})\, p\, \boldsymbol{u},
\label{aids_c} \\[6pt]
\partial_{t_1} (\rho E\, \boldsymbol{u})
+ \nabla_1 \cdot [(p + \rho E)\, \boldsymbol{u}\boldsymbol{u}]
&= - E\, \nabla_1 p
   + p \boldsymbol{u} \cdot \nabla_1 \boldsymbol{u}. \label{aids_d}
\end{align}
\end{subequations}

Here, $[\cdot]_s$ represents the symmetric part of the second-order tensor in the brackets.
For instance,
$[\boldsymbol{u} \nabla_1 p]_s = \boldsymbol{u} \nabla_1 p + (\boldsymbol{u} \nabla_1 p)^T$,
with the superscript $T$ denoting the tensor transpose.
By using Eqs.~\eqref{moments_b}, \eqref{aids_a}, and \eqref{aids_b}, we can obtain
\begin{align}
-\frac{1}{\tau_f}\boldsymbol{\Pi}^{(1)}
&= \partial_{t_1} \int \boldsymbol{\xi}\,\boldsymbol{f}^{(0)} d\boldsymbol{\xi}
   + \nabla_1 \cdot 
     \int \boldsymbol{\xi}\boldsymbol{\xi}\,\boldsymbol{f}^{(0)} d\boldsymbol{\xi} \notag \\
&= \partial_{t_1} p\, \boldsymbol{I}
 + \partial_{t_1} (\rho\, \boldsymbol{u} \boldsymbol{u})
 + [\nabla_1 (\rho \boldsymbol{u})]_{s}
 + \nabla_1 \cdot (p\, \boldsymbol{u} \boldsymbol{I})
 + \nabla_1 \cdot (\rho\, \boldsymbol{u}\boldsymbol{u}\boldsymbol{u}) \notag \\
&= p\!\left(
     [\nabla_1 \boldsymbol{u}]_{s} 
     - \frac{2}{K+3}(\nabla_1 \cdot \boldsymbol{u}) \boldsymbol{I}
   \right).
\end{align}

Therefore, the viscous stress tensor can be obtained as
\begin{align}
    \boldsymbol{\sigma} &= - \boldsymbol{\Pi}^{(1)} 
    = - \int \boldsymbol{\xi} \boldsymbol{f}^{(1)}\, d\boldsymbol{\xi} \notag \\
    &=2\mu \!\left[ \frac{1}{2}[\nabla_1 \boldsymbol{u}]_{s}
        - \frac{1}{D} (\nabla_1\!\cdot\!\boldsymbol{u})\, \boldsymbol{I} \right]
        + \mu_b (\nabla_1\!\cdot\!\boldsymbol{u})\, \boldsymbol{I},
\end{align}
where $\mu=\tau_f p$ is the shear viscosity and $\mu_b = 2\mu(K+3-D)/D(K+3)$ is the bulk viscosity.

Similarly, by using Eqs.~\eqref{moments_c}, \eqref{aids_c}, and \eqref{aids_d},
\begin{align}
-\frac{1}{\tau_e}\boldsymbol{Q}^{(1)}
+ \frac{1}{\tau_{ef}}\int \boldsymbol{\xi}\, R^{(1)} d\boldsymbol{\xi}
&= \partial_{t_1}\int \boldsymbol{\xi}\, f_{D+1}^{(0)} d\boldsymbol{\xi}
 + \nabla_1\!\cdot\! \int \boldsymbol{\xi}\, \boldsymbol{\xi}\, f_{D+1}^{(0)} d\boldsymbol{\xi}
\notag \\
&= \partial_{t_1}(p\boldsymbol{u})
 + \partial_{t_1}(\rho E \boldsymbol{u})
 + \nabla_1 [\, p(RT + E)\,]
 + \nabla_1\!\cdot(p\,\boldsymbol{u}\boldsymbol{u})
 + \nabla_1\!\cdot[(p+\rho E)\,\boldsymbol{u}\boldsymbol{u}]
\notag \\
&= p\!\left(
      R\nabla_1 T
      - \frac{2}{K+3}\boldsymbol{u} (\nabla_1\!\cdot\!\boldsymbol{u})
      + \boldsymbol{u}\cdot\nabla_1 \boldsymbol{u}
      + \nabla_1 E
   \right)
\notag \\
&= \frac{K+5}{2}\, pR\nabla_1 T
 + p\boldsymbol{u}\cdot\Bigl(
       [\nabla_1 \boldsymbol{u}]_{s} 
     - \frac{2}{K+3}(\nabla_1\!\cdot\!\boldsymbol{u})\,\boldsymbol{I}
   \Bigr)
\notag \\
&= \frac{\lambda}{\tau_e} \nabla_1 T 
   + \frac{1}{\tau_f} \boldsymbol{\sigma} \cdot \boldsymbol{u}.
\end{align}

Since
\begin{align}
    \int \boldsymbol{\xi}\, R^{(1)} \, d\boldsymbol{\xi} 
    &= \int \boldsymbol{\xi}\,
        \left( \boldsymbol{f} - \boldsymbol{f}^{(eq)} \right) \cdot \boldsymbol{u}
       \, d\boldsymbol{\xi} \notag \\
    &\approx \int \boldsymbol{\xi}\, \boldsymbol{f}^{(1)} \cdot \boldsymbol{u}
       \, d\boldsymbol{\xi}
    = - \boldsymbol{\sigma} \cdot \boldsymbol{u},
\end{align}
and we can obtain
\begin{align}
   - \boldsymbol{Q}^{(1)}
   = \lambda \nabla_1 T
   + \boldsymbol{\sigma} \cdot \boldsymbol{u},
\end{align}
where $\lambda = (K+5) R \tau_e p / 2$ denotes the thermal conductivity.

\newpage
\section{Discrete velocity sets}
\setcounter{table}{0} \renewcommand{\thetable}{B.\arabic{table}}
In this appendix, we list several typical DVSs satisfying the requirements for recovering the compressible Navier-Stokes equations, as shown in Tab.~\ref{tab:DVS}. Note that two choices are available for the D3V27-A7 set, and the set with the upper signs is denoted as Group I, and the other as Group II.
\begin{table}[htbp]
\centering
\renewcommand{\arraystretch}{1.085} % 增加行高
\setlength{\tabcolsep}{16pt} % 增加列间距
\caption{\small The abscissae $\hat{\boldsymbol{\xi}}_i$ and corresponding weights $w_i$ for several 3D and 2D DVSs. The term “cycle” denotes a cyclic permutation. The notation D$n$V$b$--A$k$ refers to a  DVS containing $b$ discrete velocities in a $D$-dimensional space, with a quadrature precision of degree $k$. The data are compiled from \cite{SHAN_YUAN_CHEN_2006}.}
\begin{tabular}{lccc}
\toprule
\textbf{DVS} & $\hat{\boldsymbol{\xi}}_i$ & $w_i$ &  \\
\midrule
D1V3-A5 & $0$ & $2/3$ & $s = \sqrt{3}$\\
        & $\pm s$ & $1/6$ & \\
D1V4-A7 & $\pm r$ & ${(3 + \sqrt{6})}/{12}$ & $3 = \sqrt{3 - \sqrt{6}}$\\
        & $\pm s$ & ${(3 - \sqrt{6})}/{12}$ & $s = \sqrt{3 + \sqrt{6}}$\\
D2V6-A4 & $(0, 0)$ & $1/2$ & $\theta_k = 2k\pi/5,\ k = 1 \sim 5$ \\
        & $2(\cos\theta_k, \sin\theta_k)$ & $1/10$ & \\
D2V7-A5 & $(0, 0)$ & $1/2$ & $\theta_k = k\pi/3,\ k = 1 \sim 6$ \\
        & $2(\cos\theta_k, \sin\theta_k)$ & $1/12$ & \\
D2V12-A7 & $(\pm r, 0)$ (cycle) & $1/36$ & $r = \sqrt{6}$ \\
         & $(\pm s, \pm s)$ & $(5 + 2\sqrt{5})/45$ & $s = \sqrt{9 - 3\sqrt{5}}/2$ \\
         & $(\pm q, \pm q)$ & $(5 - 2\sqrt{5})/45$ & $q = \sqrt{9 + 3\sqrt{5}}/2$ \\
D3V13-A5 & $(0, 0, 0)$ & $2/5$ & $r = \sqrt{(5 + \sqrt{5})/2}$ \\
         & $(\pm r, \pm s, 0)$ (cycle) & $1/20$ & $s = \sqrt{(5 - \sqrt{5})/2}$ \\
D3V27-A7 & $(0, 0, 0)$ & $(720 \pm 8\sqrt{15})/{2205}$ & \\
         & $(\pm r, 0, 0)$ (cycle) & ${(270 \mp 46\sqrt{15})}/{15435}$ & $r = \sqrt{(15 \pm \sqrt{15})/2}$ \\
         & $(\pm s, \pm s, 0)$ (cycle) & $({162 \pm 41\sqrt{15}})/{6174}$ & $s = \sqrt{6 \mp \sqrt{15}}$ \\
         & $(\pm q, \pm q, \pm q)$ & $({783 \mp 202\sqrt{15}})/{24696}$ & $q = \sqrt{9 \pm 2\sqrt{15}}$ \\
\bottomrule
\end{tabular}
\label{tab:DVS}
\end{table}

\newpage
\section*{Acknowledgments}
This study was financially supported by the National Natural Science Foundation of China (Grant No. 12472290).
% Z. Xin gratefully acknowledges the support by the China Scholarship Council (No. 202406160103).
% The computation is completed in the HPC Platform of Huazhong University of Science and Technology.


%Lorem ipsum dolor sit amet, consectetur adipiscing elit, sed do eiusmod tempor section .
\bibliographystyle{elsarticle-num} 
\bibliography{myReferences}

@article{SHAN_YUAN_CHEN_2006, 
title={Kinetic theory representation of hydrodynamics: a way beyond the {Navier–Stokes} equation}, 
volume={550}, 
journal={Journal of Fluid Mechanics}, 
author={SHAN, XIAOWEN and YUAN, XUE-FENG and CHEN, HUDONG}, 
year={2006}, 
pages={413–441}
}

@article{BGKmodel1993,
title = {Numerical Hydrodynamics from Gas-Kinetic Theory},
journal = {Journal of Computational Physics},
volume = {109},
number = {1},
pages = {53-66},
year = {1993},
issn = {0021-9991},
author = {Kevin H. Prendergast and Kun Xu}
}

@misc{chapman1970,
  title={The mathematical theory of non uniform gases, cambridge mathematical library},
  author={Chapman, Sydney and Cowling, TG},
  year={1970},
  publisher={Cambridge Univ. Press, Cambridge, UK}
}

@article{GUO2023Compressible,
  title = {Discrete unified gas kinetic scheme for continuum compressible flows},
  author = {Guo, Zhaoli and Wang, Lian-Ping and Qi, Yiming},
  journal = {Phys. Rev. E},
  volume = {107},
  issue = {2},
  pages = {025304},
  numpages = {14},
  year = {2023},
  month = {Feb},
  publisher = {American Physical Society}
}

@article{VANLEER1977,
title = {Towards the ultimate conservative difference scheme. IV. A new approach to numerical convection},
journal = {Journal of Computational Physics},
volume = {23},
number = {3},
pages = {276-299},
year = {1977},
issn = {0021-9991},
author = {Bram {Van Leer}},
}

@article{Riemann1998,
author = {Lax, Peter D. and Liu, Xu-Dong},
title = {Solution of Two-Dimensional {Riemann} Problems of Gas Dynamics by Positive Schemes},
journal = {SIAM Journal on Scientific Computing},
volume = {19},
number = {2},
pages = {319-340},
year = {1998},
}

@article{Guo2013DUGKS,
  title = {Discrete unified gas kinetic scheme for all {Knudsen} number flows: Low-speed isothermal case},
  author = {Guo, Zhaoli and Xu, Kun and Wang, Ruijie},
  journal = {Phys. Rev. E},
  volume = {88},
  issue = {3},
  pages = {033305},
  numpages = {11},
  year = {2013},
  month = {Sep},
  publisher = {American Physical Society},
}

@article{Guo2015DUGKS,
  title = {Discrete unified gas kinetic scheme for all {Knudsen} number flows. II. Thermal compressible case},
  author = {Guo, Zhaoli and Wang, Ruijie and Xu, Kun},
  journal = {Phys. Rev. E},
  volume = {91},
  issue = {3},
  pages = {033313},
  numpages = {15},
  year = {2015},
  month = {Mar},
  publisher = {American Physical Society},
}

@article{SOD19781,
title = {A survey of several finite difference methods for systems of nonlinear hyperbolic conservation laws},
journal = {Journal of Computational Physics},
volume = {27},
number = {1},
pages = {1-31},
year = {1978},
issn = {0021-9991},
author = {Gary A Sod}
}

@article{Lax1954WeakSO,
  title={Weak solutions of nonlinear hyperbolic equations and their numerical computation},
  author={Peter D. Lax},
  journal={Communications on Pure and Applied Mathematics},
  year={1954},
  volume={7},
  pages={159-193},
}

@article{SHUOsher1988439,
title = {Efficient implementation of essentially non-oscillatory shock-capturing schemes},
journal = {Journal of Computational Physics},
volume = {77},
number = {2},
pages = {439-471},
year = {1988},
issn = {0021-9991},
author = {Chi-Wang Shu and Stanley Osher},
}

@article{1937TGV,
    author = {Taylor, Geoffrey Ingram and Green, Albert Edward},
    title = {Mechanism of the production of small eddies from large ones},
    journal = {Proceedings of the Royal Society of London. A. Mathematical and Physical Sciences},
    volume = {158},
    number = {895},
    pages = {499-521},
    year = {1937},
    month = {02},
    issn = {0080-4630},
}

@article{Sutherland2021,
author = {Lusher, David J. and Sandham, Neil D.},
title = {Assessment of Low-Dissipative Shock-Capturing Schemes for the Compressible {Taylor–Green} Vortex},
journal = {AIAA Journal},
volume = {59},
number = {2},
pages = {533-545},
year = {2021},
}

@article{Sutherland2018,
  title = {Effects of the {Mach} number on the evolution of vortex-surface fields in compressible {Taylor-Green} flows},
  author = {Peng, Naifu and Yang, Yue},
  journal = {Phys. Rev. Fluids},
  volume = {3},
  issue = {1},
  pages = {013401},
  numpages = {21},
  year = {2018},
  month = {Jan},
  publisher = {American Physical Society},
}

@article{Wang2013,
author = {Wang, Z.J. and Fidkowski, Krzysztof and Abgrall, Rémi and Bassi, Francesco and Caraeni, Doru and Cary, Andrew and Deconinck, Herman and Hartmann, Ralf and Hillewaert, Koen and Huynh, H.T. and Kroll, Norbert and May, Georg and Persson, Per-Olof and van Leer, Bram and Visbal, Miguel},
title = {High-order {CFD} methods: current status and perspective},
journal = {International Journal for Numerical Methods in Fluids},
volume = {72},
number = {8},
pages = {811-845},
year = {2013}
}

@article{Wissocq2025,
author = {Wissocq, Gauthier and Liu, Yongle and Abgrall, R\'{e}mi},
title = {A Positive- And Bound-Preserving Vectorial Lattice {Boltzmann} Method in Two Dimensions},
journal = {SIAM Journal on Scientific Computing},
volume = {47},
number = {6},
pages = {A3276-A3302},
year = {2025},
}

@article{BIANCHINI2019280,
title = {Strong convergence of a vector-{BGK} model to the incompressible {Navier-Stokes} equations via the relative entropy method},
journal = {Journal de Mathématiques Pures et Appliquées},
volume = {132},
pages = {280-307},
year = {2019},
issn = {0021-7824},
author = {Roberta Bianchini}
}

@article{zhaojin2021,
  title={Discrete-velocity vector-{BGK} models based numerical methods for the incompressible {Navier-Stokes} equations},
  author={Zhao, Jin},
  journal={Communications in Computational Physics},
  volume={29},
  number={2},
  pages={420--444},
  year={2021}
}

@article{Zhao2020,
author = {Zhao, Jin and Zhang, Zhimin and Yong, Wen-An},
title = {Vector-Type Boundary Schemes for the Lattice {Boltzmann} Method Based on Vector-{BGK} Models},
journal = {SIAM Journal on Scientific Computing},
volume = {42},
number = {5},
pages = {B1250-B1270},
year = {2020},
}

@article{ZHAO2024115677,
title = {Vectorial finite-difference-based lattice {Boltzmann} method: Consistency, boundary schemes and stability analysis},
journal = {Journal of Computational and Applied Mathematics},
volume = {441},
pages = {115677},
year = {2024},
issn = {0377-0427},
author = {Jin Zhao and Wen-An Yong}
}

@article{ZHAO20202764,
title = {Seven-velocity three-dimensional vectorial lattice {Boltzmann} method including various types of approximations to the pressure and two-parameterized second-order boundary treatments},
journal = {Computers \& Mathematics with Applications},
volume = {80},
number = {12},
pages = {2764-2779},
year = {2020},
issn = {0898-1221},
author = {Jin Zhao and Zhimin Zhang}
}

@article{ZHAO2020123901,
title = {Approximation of the multi-dimensional incompressible {Navier-Stokes} equations by discrete-velocity vector-{BGK} models},
journal = {Journal of Mathematical Analysis and Applications},
volume = {486},
number = {2},
pages = {123901},
year = {2020},
issn = {0022-247X},
author = {Jin Zhao and Zhimin Zhang and Wen-An Yong}
}

@article{BATY2023127667,
title = {A robust and efficient solver based on kinetic schemes for Magnetohydrodynamics ({MHD}) equations},
journal = {Applied Mathematics and Computation},
volume = {440},
pages = {127667},
year = {2023},
issn = {0096-3003},
author = {Hubert Baty and Florence Drui and Philippe Helluy and Emmanuel Franck and Christian Klingenberg and Lukas Thanhäuser},
}

@article{caetano2019result,
  title={A result of convergence for a mono-dimensional two-velocities lattice {Boltzmann} scheme},
  author={Caetano, Filipa and Dubois, Fran{\c{c}}ois and Graille, Benjamin},
  journal={arXiv preprint arXiv:1905.12393},
  year={2019}
}

@article{Chai_2022,
   title={Multiple-distribution-function lattice {Boltzmann} method for convection-diffusion-system-based incompressible {Navier-Stokes} equations},
   volume={106},
   ISSN={2470-0053},
   number={5},
   journal={Physical Review E},
   publisher={American Physical Society (APS)},
   author={Chai, Zhenhua and Shi, Baochang and Zhan, Chengjie},
   year={2022},
   month=nov 
}

@article{Coulette2020,
  TITLE = {{Vectorial kinetic relaxation model with central velocity. Application to implicit relaxations schemes}},
  AUTHOR = {Coulette, David and Court{\`e}s, Cl{\'e}mentine and Franck, Emmanuel and Navoret, Laurent},
  JOURNAL = {{Communications in Computational Physics}},
  PUBLISHER = {{Global Science Press}},
  VOLUME = {27},
  NUMBER = {4},
  YEAR = {2020},
  HAL_ID = {hal-01942317},
  HAL_VERSION = {v1},
}

@article{Dubois2014,
author = {Dubois, Fran\c{c}ois},
title = {Simulation of strong nonlinear waves with vectorial lattice {Boltzmann} schemes},
journal = {International Journal of Modern Physics C},
volume = {25},
number = {12},
pages = {1441014},
year = {2014},
}

@article{GRAILLE201474,
title = {Approximation of mono-dimensional hyperbolic systems: A lattice {Boltzmann} scheme as a relaxation method},
journal = {Journal of Computational Physics},
volume = {266},
pages = {74-88},
year = {2014},
issn = {0021-9991},
author = {B. Graille}
}

@article{bouchut1999construction,
  title={Construction of {BGK} models with a family of kinetic entropies for a given system of conservation laws},
  author={Bouchut, Fran{\c{c}}ois},
  journal={Journal of Statistical Physics},
  volume={95},
  number={1},
  pages={113--170},
  year={1999},
  publisher={Springer}
}

@article{DELLAR200295,
title = {Lattice Kinetic Schemes for Magnetohydrodynamics},
journal = {Journal of Computational Physics},
volume = {179},
number = {1},
pages = {95-126},
year = {2002},
issn = {0021-9991},
author = {Paul J. Dellar}
}

@article{bouchut2003entropy,
  title={Entropy satisfying flux vector splittings and kinetic {BGK} models},
  author={Bouchut, Fran{\c{c}}ois},
  journal={Numerische Mathematik},
  volume={94},
  number={4},
  pages={623--672},
  year={2003},
  publisher={Springer}
}

@misc{Roberta2019,
      title={Convergence of a vector {BGK} approximation for the incompressible {Navier-Stokes} equations}, 
      author={Roberta Bianchini and Roberto Natalini},
      year={2019},
      archivePrefix={arXiv},
      primaryClass={math.AP},
}

@article{Guillon2024,
	author = {{Guillon, Kévin} and {Hélie, Romane} and {Helluy, Philippe}},
	title = {Stability analysis of the vectorial Lattice-{Boltzmann} method},
	journal = {ESAIM: ProcS},
	year = 2024,
	volume = 77,
	pages = "46-78",
}

@article{BGK1954,
  title = {A Model for Collision Processes in {Gases. I. Small} {Amplitude} {Processes} in {Charged and Neutral One-Component Systems}},
  author = {Bhatnagar, P. L. and Gross, E. P. and Krook, M.},
  journal = {Phys. Rev.},
  volume = {94},
  issue = {3},
  pages = {511--525},
  numpages = {0},
  year = {1954},
  month = {May},
  publisher = {American Physical Society},
}

@article{WILDE2021101355,
title = {Cubature rules for weakly and fully compressible off-lattice {Boltzmann} methods},
journal = {Journal of Computational Science},
volume = {51},
pages = {101355},
year = {2021},
issn = {1877-7503},
author = {Dominik Wilde and Andreas Krämer and Mario Bedrunka and Dirk Reith and Holger Foysi}
}

@article{Wenxin2021,
    author = {Wen, Xin and Wang, Lian-Ping and Guo, Zhaoli },
    title = {Designing a consistent implementation of the discrete unified gas-kinetic scheme for the simulation of three-dimensional compressible natural convection},
    journal = {Physics of Fluids},
    volume = {33},
    number = {4},
    pages = {046101},
    year = {2021},
    month = {04},
    issn = {1070-6631},
}

@article{qiyiming2022,
    author = {Qi, Yiming and Chen, Tao and Wang, Lian-Ping and Guo, Zhaoli and Chen, Shiyi },
    title = {An efficient discrete unified gas-kinetic scheme for compressible turbulence},
    journal = {Physics of Fluids},
    volume = {34},
    number = {11},
    pages = {116101},
    year = {2022},
    month = {11},
    issn = {1070-6631},
}

@article{qi2023regularized,
  title = {Consistent lifting relations for the initialization of total-energy double-distribution-function kinetic models},
  author = {Qi, Yiming and Wang, Lian-Ping and Guo, Zhaoli and Chen, Shiyi},
  journal = {Phys. Rev. E},
  volume = {108},
  issue = {6},
  pages = {065301},
  numpages = {22},
  year = {2023},
  month = {Dec},
  publisher = {American Physical Society},
}

@inproceedings{qi2024regularized,
  title={A regularized discrete unified gas-kinetic scheme for incompressible viscous flows},
  author={Qi, Yiming and Shen, Jie and Guo, Zhaoli and Chen, Shiyi and Wang, Lian-Ping},
  booktitle={APS Division of Fluid Dynamics Meeting Abstracts},
  pages={S01--032},
  year={2024}
}

@article{Dominik2021semi,
  title = {High-order semi-Lagrangian kinetic scheme for compressible turbulence},
  author = {Wilde, Dominik and Kr\"amer, Andreas and Reith, Dirk and Foysi, Holger},
  journal = {Phys. Rev. E},
  volume = {104},
  issue = {2},
  pages = {025301},
  numpages = {15},
  year = {2021},
  month = {Aug},
  publisher = {American Physical Society},
}

@article{guo2026unified,
  title={A unified gas-kinetic framework from {Boltzmann} to {Navier-Stokes} scales},
  author={Guo, Zhaoli and Xu, Kun and Zhu, Yajun},
  journal={Advances in Aerodynamics},
  volume={8},
  number={1},
  pages={8},
  year={2026},
  publisher={Springer}
}

@article{bellotti2025fourth,
  title={Fourth-order entropy-stable lattice {Boltzmann} schemes for hyperbolic systems},
  author={Bellotti, Thomas and Helluy, Philippe and Navoret, Laurent},
  journal={SIAM Journal on Scientific Computing},
  volume={47},
  number={1},
  pages={A586--A611},
  year={2025},
  publisher={SIAM}
}





% \end{thebibliography}
\end{document}